\documentclass[reqno,12pt]{amsart}
\usepackage{amssymb,amsmath}
\usepackage{bbm}
\usepackage{fullpage}

\begin{document}
 
\newcommand{\vspacebefore}{\raisebox{0ex}[2.5ex][0ex]{\null}}
\newcommand{\vspacebeforemore}{\raisebox{0ex}[3.5ex][0ex]{\null}}
\newcommand{\p}{\partial}
\newcommand{\const}{\mathop{\rm const}\nolimits}
\newcommand{\Equiv}{\mathop{\rm \, equiv}}
\newcommand{\sign}{\mathop{\rm sign}\nolimits}
\newcommand{\RR}{\mathbbm{R}}
\newcommand{\CC}{\mathbbm{C}}
\newcommand{\di}{\partial}
\newcommand{\grad}{\nabla}
\newcommand{\wt}{\widetilde}
\newcommand{\wh}{\widehat}
\newcommand{\bea}{\begin{eqnarray*}}
\newcommand{\eea}{\end{eqnarray*}}
\newcommand{\ra}{\rightarrow}
\newcommand{\gradS}{\grad_{S^{n-1}}}
\newcommand{\laplacianS}{\triangle_{S^{n-1}}}  
\newcommand{\X}{\wt{X}}
\newcommand{\PP}{\wt{P}}
\newcommand{\Q}{\wt{Q}}
\newcommand{\T}{\wt{T}}
\newcommand{\W}{\wt{W}}
\newcommand{\A}{\wt{A}}
\newcommand{\B}{\wt{B}}
\newcommand{\LL}{\wt{L}}
\renewcommand{\i}{{\rm i}}
\newcommand{\rmX}[1]{{\rm X}_{#1}}
\newcommand{\hatrmX}[1]{\hat{\rm X}{}_{#1}}
\newcommand{\h}{\wt{h}}
\newcommand{\pr}{{\rm pr}}
\newcounter{tbn}
\newcounter{tabul}

\newtheorem{prop}{Proposition}
\newtheorem{thm}{Theorem}
\newtheorem{cor}{Corollary}

\newenvironment{changemargin}[2]{%
  \begin{list}{}{%
    \setlength{\topsep}{0pt}%
    \setlength{\leftmargin}{#1}%
    \setlength{\rightmargin}{#2}%
    \setlength{\listparindent}{\parindent}%
    \setlength{\itemindent}{\parindent}%
    \setlength{\parsep}{\parskip}%
  }%
  \item[]}{\end{list}}

\title{Conservation laws and symmetries of\\ quasilinear radial wave equations in multi-dimensions}

\author{Stephen C. Anco}
\address{Department of Mathematics, Brock University, St. Catharines, ON, L2S 3A1, Canada}

\author{Steven A. MacNaughton}
\address{Department of Mathematics, Brock University, St. Catharines, ON, L2S 3A1, Canada}

\author{Thomas Wolf}
\address{Department of Mathematics, Brock University, St. Catharines, ON, L2S 3A1, Canada}

\begin{abstract}
Symmetries and conservation laws are studied for two classes 
of physically and analytically interesting radial wave equations 
with power nonlinearities in multi-dimensions.
The results consist of two main classifications:
all symmetries of point type and all conservation laws of a general energy-momentum type
are explicitly determined,
including those such as dilations, inversions, similarity energies and conformal energies
that exist only for special powers or dimensions.
In particular, all variational cases (when a Lagrangian formulation exists) 
and non-variational cases (when no Lagrangian exists)
for these wave equations are considered.
As main results, the classification yields generalized energies and radial momenta
in certain non-variational cases, 
which are shown to arise from a new type of Morawetz dilation identity 
that produces conservation laws for each of the two classes of wave equations
in a different way than Noether's theorem. 
\end{abstract}



\maketitle

\section{Introduction}\label{sec:Introduction}

Symmetries and conservation laws are important tools in the study of global analysis of nonlinear wave equations 
\begin{equation}
u_{tt}=g(x,u,\grad u)\cdot(\grad^{2} u)+f(x,u,\grad u)
\label{class1}
\end{equation}
for $u(t,x)$ in $n\ge1$ spatial dimensions 
(where $x$ denotes Cartesian coordinates). 

As pointed out in \cite{Str,AncIva}, 
conservation laws such as energy provide basic conserved quantities
used in obtaining estimates on $|u|$ or $|\nabla u|$
for classical solutions,
and also in defining suitable norms for weak solutions.
Of considerable interest are extra conservation laws
such as conformal energies that can appear in the case of 
power nonlinearities $|u|^p$ or derivative nonlinearities $|\grad u|^p$ 
in $g$ and $f$ for special powers $p$ depending on the dimension $n$.
Symmetries, in contrast, lead to exact group-invariant solutions
and play a role in defining invariant Sobolev norms.
Scaling symmetries are of special relevance,
as the critical nonlinearity power for a blow-up is typically singled out by
scaling-invariance of a positive energy norm.
Moreover, scaling transformation arguments give a means of
relating the behavior of solutions
in different regimes, for instance,
solutions at short times with large initial data
can be scaled to long times with small initial data
when the nonlinearity power is subcritical.

In previous work by Anco\&Ivanova \cite{AncIva}, 
conservation laws and symmetries have been classified for 
semilinear wave equations
\begin{equation}
 u_{tt}=\triangle u+au^{p}
\end{equation}
with a power nonlinearity in $n>1$ spatial dimensions. 
In the present paper and a sequel, 
we extend those classifications to a variety of physically and analytically interesting quasilinear wave equations 
\begin{subequations}\label{WEequiv}
\begin{equation}\label{WEequiv1}
 u_{tt}=\grad\cdot\Big(\big(c+\dfrac{a}{p}u^{p}\big)\grad u\Big)+\big(b-\dfrac{a}{p}\big)u^{p}\triangle u
\end{equation}
and
\begin{equation}\label{WEequiv2}
 u_{tt}=\grad\cdot\Big(\big(c+\dfrac{a}{p}|\grad u|^{p}\big)\grad u\Big)+\big(b-\dfrac{a}{p}\big)|\grad u|^{p}\triangle u
\end{equation}
\end{subequations}
contained in the general class~\eqref{class1}, 
with free constants $a$, $b$, $c$, 
where $p\neq 0$ is the nonlinearity power and $n>1$ is the spatial dimension. 
These wave equations can be written respectively in the equivalent forms
\begin{subequations}\label{WE}
\begin{equation}
u_{tt}=(c+bu^{p})\triangle u+au^{p-1}|\grad u|^{2}
\label{WE1}
\end{equation}
and
\begin{equation}
u_{tt}=(c+b|\grad u|^{p})\triangle u +\tfrac{1}{2}a|\grad u|^{p-2}(\grad u\cdot\grad)^{2}u
\label{WE2}
\end{equation}
\end{subequations}
which are valid for any power $p$. 
There has been considerable mathematical interest in the analysis 
(e.g. global existence, uniqueness, regularity, blow up) of solutions to 
the Cauchy problem for translationally-invariant quasilinear wave equations
in multi-dimensions \cite{Ali,Sog}, 
particularly for radially symmetric initial data. 
The two classes of wave equations \eqref{WE1}--\eqref{WE2} studied here
provide the simplest such examples having homogeneous power nonlinearities. 

To begin, we will consider 
the radial reduction of the wave equations \eqref{WE1} and \eqref{WE2} for $u(t,r)$, where $r=|x|$ is the radial
coordinate in $\RR^{n}$. 
This reduction yields
\begin{subequations}\label{WEreducrad}
\begin{equation}\label{WEreducrad1}
u_{tt}=(c+bu^{p})\big(u_{rr}+\dfrac{n-1}{r}u_{r}\big)+au^{p-1}u_{r}^{2}
\end{equation}
and
\begin{equation}\label{WEreducrad2}
u_{tt}=(c+(b+a)u_{r}^{p})\big(u_{rr}+\dfrac{n-1}{r}u_{r}\big)-a\dfrac{n-1}{r}u_{r}^{p+1}
\end{equation}
\end{subequations}
respectively, which are radial wave equations belonging to the general class
\begin{equation}\label{WErad}
u_{tt}=g(r,u,u_{r})u_{rr}+f(r,u,u_{r})
\end{equation}
where
\begin{subequations}
\begin{equation}\label{WErad1}
g=c+bu^{p},\quad f=\dfrac{n-1}{r}(c+bu^{p})u_{r}+au^{p-1}u_{r}^{2}
\end{equation}
for equation \eqref{WEreducrad1}, 
and where
\begin{equation}\label{WErad2}
g=c+(b+a)u_{r}^{p},\quad f=\dfrac{n-1}{r}(cu_{r}+bu_{r}^{p+1})
\end{equation}
\end{subequations}
for equation \eqref{WEreducrad2}. 
We note that the class \eqref{WErad1} 
is quasilinear hyperbolic iff\ $bp\neq0$ 
or is semilinear hyperbolic iff\ $bp=0$, $a\neq0$ and $c+b\neq0$; 
and the class \eqref{WErad2} 
is quasilinear hyperbolic iff\ $(b+a)p\neq0$ 
or is semilinear hyperbolic iff\ $b+a=0$, $bc\neq0$ and $p(p+1)\neq0$.

These classes \eqref{WErad1} and \eqref{WErad2} of radial wave equations 
describe models for nonlinear radial wave propagation in which the wave speed
can depend on the wave amplitude or its radial gradient. 
Such models encompass a wide variety of physically interesting examples 
in $n=2$ and $n=3$ dimensions:
for class \eqref{WErad1}, 
radial compressible polytropic fluid flow 
in Lagrangian coordiantes \cite{MeiPukShm}
and radial propagation of voltage/current 
in electromagnetic transmission guides \cite{Kon}; 
and for class \eqref{WErad1}, 
radial nonlinear vibrations of membranes 
and radial deformations of hyperelastic materials
\cite{FuOgd,LebClo}. 
The extension of these examples to arbitrary $n>1$ dimensions is 
mathematically natural in light of the analytical interest in $n$-dimensional
quasilinear wave equations. 
Consequently, our results in this paper should be of direct relevance to
investigation of physical models of nonlinear radial wave propagation 
as well as to on-going analytical work on the Cauchy problem for radial quasilinear wave equations. 

There are two important cases in which the radial wave equations 
\eqref{WErad1} and \eqref{WErad2} have a special analytical structure. 
First, 
a radial wave equation \eqref{WErad} has a $n$-dimensional divergence structure if 
its derivative terms can be expressed as a total divergence of the form
\begin{equation}\label{divergence}
 (r^{n-1}u_{t})_{t}=(r^{n-1}F)_{r}
\end{equation}
for some function $F(u,u_{r})$. 
Such equations \eqref{divergence} give rise to kinematic conservation laws involving the wave amplitude $u$ and its time derivative $u_{t}$, which will be discussed later. 
Second, 
a radial wave equation \eqref{WErad} has a $n$-dimensional variational structure if 
it arises as the stationary points of a functional 
\begin{equation}\label{variational}
 \mathcal{L}=\int_{t_0}^{t_1}\int_{0}^{\infty}Lr^{n-1} drdt, \quad 
\frac{\delta \mathcal{L}}{\delta u}=r^{n-1}(u_{tt}-gu_{rr}-f)=0
\end{equation}
for some Lagrangian $L(u,u_{t},u_{r})$. 
Wave equations with this structure \eqref{variational} possess an energy conservation law related to time-translation symmetry through Noether's theorem \cite{Olv,2ndbook}, 
as will be reviewed in more detail later.

Wave equation~\eqref{WEreducrad1} is variational when (and only when) its coefficients $a$, $b$, $c$ satisfy the relation $2a=pb$, 
where the Lagrangian is given by
\begin{subequations}\label{Lagrang}
\begin{equation}\label{Lagrang1}
 L=\tfrac{1}{2}(-u_{t}^{2}+(c+bu^{p})u_{r}^{2}).
\end{equation}
Wave equation~\eqref{WEreducrad2} is variational when (and only when) its coefficients $a$, $b$, $c$ satisfy the relation $a=pb$. 
The Lagrangian is given by
\begin{equation}\label{Lagrang2}
 L=\tfrac{1}{2}(-u_{t}^{2}+cu_{r}^{2})+bh(u_{r})
\end{equation}
\end{subequations}
where
\begin{equation}
 h=\int u_{r}^{p+1}du_{r}=\left\{ \begin{aligned} &\dfrac{1}{p+2}u_{r}^{p+2},\ &p\neq-2\\& \ln{u_{r}},\ &p=-2\end{aligned} \right. .
\end{equation}

Both wave equations \eqref{WEreducrad1} and \eqref{WEreducrad2} are total divergences when (and only when) $a=bp$, 
with the respective fluxes given by
\begin{subequations}\label{fluxes}
 \begin{equation}\label{flux1}
  F=(c+bu^{p})u_{r}=(cu+bH(u))_{r}
 \end{equation}
and
\begin{equation}\label{flux2}
 F=(c+bu_{r}^{p})u_{r}
\end{equation}
\end{subequations}
where
\begin{equation}\label{intcasesplit}
 H=\int u^{p}du=\left\{ \begin{aligned} &\dfrac{1}{p+1}u^{p+1},\ &p\neq-1\\&\ln{u},\ &p=-1\end{aligned} \right. .
\end{equation} 
We note that the conditions for existence of a divergence structure \eqref{divergence}
and a variational structure \eqref{variational} coincide for wave equation \eqref{WEreducrad2}.

Interestingly, 
wave equation \eqref{WEreducrad2} has an alternative divergence structure and variational structure 
when $c=0$ without any conditions on $a$ and $b$. 
In this case 
\begin{equation}\label{WEradc0}
 u_{tt}=(a+b)u_{r}^{p}u_{rr}+b\dfrac{n-1}{r}u_{r}^{p+1}
\end{equation}
is a total divergence of the form
\begin{equation}\label{altdivergence}
(r^{m}u_{t})_t=(r^{m}\wt{F})_{r},\quad 
\wt{F}=\left\{
\begin{aligned}&\dfrac{b+a}{p+1}u_{r}^{p+1},\ &p\neq-1\\ & (b+a)\ln{u_{r}}+b(n-1)\ln{r},\ &p=-1\end{aligned}
\right. 
\end{equation}
and also arises from a Lagrangian functional
\begin{equation}\label{altvariational}
 \mathcal{L}=\int_{t_0}^{t_1}\int_{0}^{\infty}\wt{L}r^{m}drdt,\quad 
 \wt{L}=-\tfrac{1}{2}u_{t}^{2}+\int \wt{F}(r,u_{r})du_{r}
\end{equation}
where
\begin{equation}\label{m}
 m=\dfrac{b(n-1)(p+1)}{a+b}.
\end{equation}

Compatibility between variational structures, divergence structures, quasilinearity or semilinearity is summarized by the following result.  

\begin{prop}\label{structures}
{\rm (i)}
A radial wave equation \eqref{WEreducrad1} has 
a hyperbolic, quasilinear, variational form \eqref{variational} iff $2a=pb\neq0$, 
or a hyperbolic, quasilinear, divergence form \eqref{divergence} iff $a=pb\neq0$.
(In particular, these two structures are mutually incompatible). 
\newline
{\rm (ii)}
A radial wave equation \eqref{WEreducrad2} has 
a hyperbolic, quasilinear, variational/divergence form \eqref{divergence}--\eqref{variational} iff $a=pb$, $p(p+1)b\neq0$, 
or a hyperbolic, quasilinear, alternative variational/divergence form \eqref{altdivergence}--\eqref{altvariational} iff $c=0$, $p(a+b)\neq0$. 
(Moreover, these structures coincide when $c=0$, $a=pb$, $p(p+1)b\neq0$, 
so thus $m=n-1$.)
\newline
{\rm (iii)}
A radial wave equation \eqref{WEreducrad1} and \eqref{WEreducrad2} cannot have
a hyperbolic, semilinear, variational or divergence form. 
\end{prop}

In this paper, 
for the two classes of radial wave equations \eqref{WErad1} and \eqref{WErad2},
the main goals of our work will be to determine:

({\rm i}) what symmetries are admitted other than time-translation;

({\rm ii}) what conservation laws are admitted other than energy in the variational case \eqref{variational}, 
and kinematic quantities in the divergence case \eqref{divergence}. 

In section \ref{sec:symm-conslaws}, 
we first recall the definitions of symmetries and conservation laws 
for the general class of radial wave equations \eqref{WErad}. 
For the case of a radial divergence structure \eqref{divergence} or \eqref{altdivergence}, 
we discuss the notion of kinematic conservation laws. 
For the case of a radial variational structure \eqref{variational} or \eqref{altvariational}, 
we give a statement of Noether's theorem and summarize the energy conservation law arising from time-translation symmetry.

In sections \ref{sec:symmresults} and \ref{sec:conslawresults}, 
we present the classifications of symmetries and conservation laws for 
each class of radial wave equations \eqref{WEreducrad1} and \eqref{WEreducrad2}.
For special nonlinearity powers $p$ and for special relations among the coefficients $a$, $b$, $c$, 
both of these wave equations are found to possess a variety of interesting symmetries,
including scalings, non-rigid radial dilations, and a temporal inversion,
as shown in section \ref{sec:symmresults}. 
The commutator algebra for all of the symmetries is worked out in detail. 
Variational symmetries are also identified, which yield conservation laws through the variational structures for each 
of the wave equations.
As shown in section \ref{sec:conslawresults}, these Noether conservation laws include similarity energies, 
conformal energies, and radial momenta.
Interestingly, additional conservation laws are found to exist for certain nonlinear powers $p$ and for certain relations among
the coefficients $a$, $b$, $c$.
These conservation laws describe generalized-energies that do not arise from Noether's theorem.
We account for them by deriving a variational type of Morawetz radial dilation identity that produces conservation
laws from symmetries in a different manner than the usual Noether correspondence.
This main result is explained in subsections \ref{sec:AMorawetz} and \ref{sec:BMorawetz} for the respective
wave equations \eqref{WEreducrad1} and \eqref{WEreducrad2}.
 
Our classification results will be extended to the original two classes of translationally-invariant wave equations \eqref{WE1}
and \eqref{WE2} in a subsequent paper.

\section{Preliminaries}\label{sec:symm-conslaws}

To proceed we state the definitions of symmetries and conservation laws from an analytical perspective (see also \cite{2ndbook}). 
A {\em point symmetry} of a wave equation~\eqref{WErad} is a group of transformations 
given by an infinitesimal generator
\begin{equation}\label{infinigen}
\delta t=\tau(t,r,u),\qquad \delta x=\xi(t,r,u), \qquad \delta u=\eta(t,r,u)
\end{equation}
on the variables $t$, $r$, $u$,
under which the wave equation~\eqref{WErad} is preserved. 
On solutions $u(t,r)$ of equation \eqref{WErad}, 
such an infinitesimal transformation \eqref{infinigen} is equivalent to a generator
\begin{equation}\label{pointsymm}
{\rm X}=(\eta-\tau u_t-\xi u_{r})\partial/\partial u
\end{equation}
called the {\em characteristic form} of the point symmetry. 
The scalar functions $\eta$, $\tau$, $\xi$ are determined by
a linear equation arising from the Fr\'echet derivative of the wave equation~\eqref{WErad} applied to
the function $P=\eta-\tau u_t-\xi u_{r}$, 
\begin{equation}\label{detsymm}
 0=D_{t}^{2}P-\Big(\frac{\partial f}{\partial u}+u_{rr}\frac{\partial g}{\partial u}\Big)P
    -\Big(\frac{\partial f}{\partial u_{r}}+u_{rr}\frac{\partial g}{\partial u_{r}}\Big)D_{r}P-gD_{r}^{2}P
\end{equation}
holding for all formal solutions $u(t,r)$, namely, with $u_{tt}$ (and its $r$-derivatives) replaced by $gu_{rr}+f$ 
(and its total $r$-derivatives).
(More precisely, symmetries are computed in a jet space whose coordinates are defined by
$t$, $r$, $u$ and all derivatives of $u$ modulo the equation~\eqref{WErad} and its differential consequences.
A point symmetry is then the prolongation of the operator~\eqref{pointsymm} that annihilates the wave 
equation~\eqref{WErad}.)

The definition of a symmetry can be extended to involve first-order derivatives of $u$ by considering 
an infinitesimal generator of the form
\begin{equation}\label{symgen}
 {\rm X}=P(t,r,u,u_{t},u_{r})\partial/\partial u
\end{equation}
where $P$ is a scalar function depending on $u_{t}$, $u_{r}$, in addition to $t$, $r$, $u$. 
Such a generator corresponds to the infinitesimal transformations \cite{2ndbook}
\begin{subequations}\label{contacttrans}
 \begin{align}
  &\delta t=-\frac{\partial P}{\partial u_{t}},\quad 
  \delta r=-\frac{\partial P}{\partial u_{r}},
 \\
  &\delta u=P- u_{t}\frac{\partial P}{\partial u_{t}} -u_{r}\frac{\partial P}{\partial u_{r}},
 \\ 
  &\delta u_{t}=\frac{\partial P}{\partial t} +u_{t}\frac{\partial P}{\partial u},\quad
 \\
  &\delta u_{r}=\frac{\partial P}{\partial r}+u_{r}\frac{\partial P}{\partial u}
 \end{align}
\end{subequations}
on the variables $r$, $t$, $u$, $u_{t}$, $u_{r}$ (viewed as coordinates in jet space). 
A group of transformations \eqref{contacttrans} is a \emph{contact symmetry} of a wave equation \eqref{WErad} if
these transformations preserve the equation \eqref{WErad}.

The set of all infinitesimal point or contact symmetries admitted by a given wave equation~\eqref{WErad} has 
the structure of a Lie algebra under commutation of the operators ${\rm X}$ (prolonged to the jet space). 
For a given (sub)algebra of point or contact symmetries,
the corresponding group of transformations has a natural action \cite{Olv,1stbook,2ndbook}
on the set of all solutions $u(t,r)$ of the wave equation.
A solution $u(t,r)$ is invariant under a one-dimensional (sub)group with a generator \eqref{symgen}
if it satisfies the equation $P(t,r,u,u_{t},u_{r})=0$ where
$P=\eta-\tau u_{t}-\xi u_{r}$ in the case of a point symmetry. 
Such solutions of a wave equation \eqref{WErad} are called {\em group invariant}.

A {\em conservation law} of a wave equation~\eqref{WErad}
is a space-time divergence given by a linear combination of
the equation and its differential consequences, so that
\begin{equation}\label{conlaw}\textstyle
D_t T+D_{r}X=0
\end{equation}
holds for all formal solutions~$u(t,r)$, where $T$ and $X$ are scalar functions depending on 
$t$, $r$, $u$, $u_{t}$, and $r$-derivatives of $u$, $u_{t}$. 
The radial integral of the conserved density $T$ formally satisfies
\begin{equation}
\frac d{dt} \int_{0}^{\infty} T dr=-X\big|_{0}^{\infty}\label{dtconlaw}
\end{equation}
and so if the radial flux $X$ vanishes both at spatial infinity and at the origin,
then 
\begin{equation}\label{C}
C= \int_{0}^{\infty} T dr=\text{const.}
\end{equation}
formally yields a conserved quantity for the wave equation~\eqref{WErad}.
Conversely, any such conserved quantity arises from
a conservation law~\eqref{conlaw}.
Two conservation laws are equivalent if their conserved densities $T$
differ by a total radial derivative $D_{r}\Theta$
on all formal solutions $u(t,r)$,
thereby giving the same conserved quantity $C$ up to boundary terms.
Correspondingly, the fluxes $X$ of two equivalent conservation laws
differ by a total time derivative $-D_{t}\Theta$ on all formal solutions
$u(t,r)$.
The set of all conservation laws (up to equivalence)
admitted by a given wave equation~\eqref{WErad}
forms a vector space,
on which there is a natural action \cite{BluTemAnc}
by the Lie group of all point or contact symmetries of the equation~\eqref{WErad}.

Each conservation law~\eqref{conlaw} has an equivalent {\em characteristic form} given by
\begin{equation}\label{characform}
D_t T+D_{r}(X-\Gamma)=(u_{tt}-gu_{rr}-f)Q
\end{equation}
with 
\begin{equation}
\Gamma= \sum_{l\geq0} \dfrac{\delta T}{\delta D_{r}^{l+1}u_{t}} D_{r}^{l}(u_{tt}-gu_{rr}-f)
\end{equation}
where $Q$ is a scalar function that depends on $t$, $r$, $u$, $u_{t}$, and $r$-derivatives of $u$, $u_{t}$. 
Such functions $Q$ whose product with equation \eqref{WErad} yields
a total space-time divergence 
(and hence a conservation law \eqref{conlaw} on all solutions $u(t,r)$) are called {\it multipliers}. 
Through the characteristic equation \eqref{characform}, 
multipliers and conserved densities (up to equivalence) 
have a one-to-one relation such that $Q$ is the variational derivative of $T$
with respect to $u_{t}$,
\begin{equation}\label{TQrelation}
 Q=\dfrac{\delta T}{\delta u_{t}}
\end{equation}
which can be inverted to express both $T$ and $X$ (modulo total derivative terms $D_{r}\Theta$ and $-D_{t}\Theta$ respectively)
in terms of $Q$ by means of an explicit integral formula  
\cite{AncBluI,AncBluII,Zha} based on a homotopy integration of the characteristic equation \eqref{characform}.

All multipliers $Q$ for a wave equation \eqref{WErad} are determined by
a linear system \cite{AncBluI,AncBluII,Olv,Zha} that holds for all formal solutions~$u(t,r)$.
(More precisely, the computation of multipliers uses the same jet space as for the computation of symmetries.)
The system consists of \cite{2ndbook} 
a linear equation given by the adjoint of the Fr\'echet derivative of equation~\eqref{WErad} applied to $Q$, 
augmented by additional linear equations formed from the Fr\'echet derivative of $Q$ itself.
Thus the determination of conservation laws via multipliers is
a kind of adjoint problem \cite{AncBlu}
of the determination of symmetries. 

From the variational relation \eqref{TQrelation}, 
conserved densities $T$ up to 1st order correspond to multipliers $Q$ of at most the same order, as given by
\begin{equation}\label{multiplierWE}\textstyle
Q(t,r,u,u_t,u_{r}) = \dfrac{\partial T(t,r,u,u_t,u_{r})}{\partial u_t}.
\end{equation}
Such multipliers are determined by the linear system
\begin{equation}\label{detcon1}
 0=D_{t}^{2}Q-\big(\frac{\partial f}{\partial u}+u_{rr}\frac{\partial g}{\partial u}\big)Q
    +D_{r}\Big(\big(\frac{\partial f}{\partial u_{r}}+u_{rr}\frac{\partial g}{\partial u_{r}}\big)Q\Big)
    -D_{r}^{2}\big(gQ\big)
\end{equation}
\begin{equation}\label{detcon2}
 0=2\frac{\partial Q}{\partial u}+D_{t}\frac{\partial Q}{\partial u_{t}}-D_{r}\frac{\partial Q}{\partial u_{r}}
\end{equation}
holding for all solutions $u(t,r)$, namely, with $u_{tt}$ (and its $r$-derivatives)
replaced by $gu_{rr}+f$ (and its total $r$-derivatives). For a given multiplier $Q$, 
the corresponding conserved density $T$ and flux $X$ can be obtained by integration of the relations
\begin{equation}\label{intmethodTX1}
 \dfrac{\partial T}{\partial u_{t}}=Q,\qquad \dfrac{\partial X}{\partial u_{r}}=-gQ,
\end{equation}
\begin{equation}\label{intmethodTX2}
 \dfrac{\partial T}{\partial u_{r}}+\dfrac{\partial X}{\partial u_{t}}=0,\qquad 
 \dfrac{\partial T}{\partial t}+\dfrac{\partial X}{\partial r}+u_{t}\dfrac{\partial T}{\partial u}
    +u_{r}\dfrac{\partial X}{\partial u}=-fQ, 
\end{equation}
arising from the characteristic equation \eqref{characform}. 
In particular, an explicit homotopy integration formula is given by
\begin{equation}\label{TfromQ}\begin{aligned}
 T(t,r,u,u_{t},u_{r})=&u_{t}\displaystyle{\int_{0}^{1}}\Big(Q(t,r,\lambda u,\lambda u_{t},\lambda u_{r})-
   \lambda u \dfrac{\partial Q}{\partial u}(t,r,\lambda u,\lambda u_{t},\lambda u_{r})\Big)d\lambda\\
  &+u_{r}\displaystyle{\int_{0}^{1}}\Big(W(t,r,\lambda u,\lambda u_{t},\lambda u_{r})+
   \lambda u \dfrac{\partial W}{\partial u}(t,r,\lambda u,\lambda u_{t},\lambda u_{r})\Big)d\lambda\\
  &-u\displaystyle{\int_{0}^{1}}\Big(f(r,\lambda u,\lambda u_{r})\dfrac{\partial Q}{\partial u_{t}}(t,r,\lambda u,\lambda u_{t},\lambda u_{r})\\
  &\qquad\quad+\dfrac{\partial Q}{\partial t}(t,r,\lambda u,\lambda u_{t},\lambda u_{r})
    -\dfrac{\partial W}{\partial r}(t,r,\lambda u,\lambda u_{t},\lambda u_{r})\Big)d\lambda
\end{aligned}
\end{equation}
and
\begin{equation}\label{XfromQ}\begin{aligned}
 X(t,r,u,u_{t},u_{r})=&-u_{t}\displaystyle{\int_{0}^{1}}\Big(W(t,r,\lambda u,\lambda u_{t},\lambda u_{r})+
   \lambda u \dfrac{\partial W}{\partial u}(t,r,\lambda u,\lambda u_{t},\lambda u_{r})\Big)d\lambda\\
  &-u_{r}\displaystyle{\int_{0}^{1}}\Big((Q g)(t,r,\lambda u,\lambda u_{t},\lambda u_{r})-
   \lambda u \dfrac{\partial (Q g)}{\partial u}(t,r,\lambda u,\lambda u_{t},\lambda u_{r})\Big)d\lambda\\
  &+u\displaystyle{\int_{0}^{1}}\Big(\dfrac{\partial Q}{\partial r}(t,r,\lambda u,\lambda u_{t},\lambda u_{r})
      g(t,r,\lambda u,\lambda u_{t},\lambda u_{r})\\
  &\qquad\quad\ -\dfrac{\partial (Qf)}{\partial u_{r}}(t,r,\lambda u,\lambda u_{t},\lambda u_{r})
    -\dfrac{\partial W}{\partial t}(t,r,\lambda u,\lambda u_{t},\lambda u_{r})\Big)d\lambda
\end{aligned}
\end{equation}
where
\begin{equation}\label{WforTX}
 W(t,r,u,u_{r},u_{t})=u_{t}\displaystyle{\int_{0}^{1}}\dfrac{\partial Q}{\partial u_{r}}(t,r,u,\lambda u_{t},\lambda u_{r})d\lambda
 +u_{r}\displaystyle{\int_{0}^{1}}g(r,u,\lambda u_{r})\dfrac{\partial Q}{\partial u_{t}}(t,r,u,\lambda u_{t},\lambda u_{r})d\lambda .
\end{equation}
These expressions \eqref{TfromQ}--\eqref{WforTX} are an improvement over the
standard homotopy integral formula in which both $T$ and $X$ would contain $r$-derivatives of $u_{t}$, $u_{r}$ that have been canceled here
by the addition of suitable total derivative terms.
(The proof that $T$ and $X$ satisfy relations \eqref{intmethodTX1}--\eqref{intmethodTX2} assumes that 
the scaled characteristic expression
\begin{equation*}
\big(\lambda u_{tt}-g(r,\lambda u,\lambda u_{r})\lambda u_{rr}-f(r,\lambda u,\lambda u_{r})\big)
  Q(t,r,\lambda u,\lambda u_{t},\lambda u_{r})
\end{equation*}
 as well as the scaled differential form
\begin{equation*}
\dfrac{\partial Q}{\partial u_{r}}(t,r,u,\lambda u_{t},\lambda u_{r})\lambda du_{t}+
  g(r,u,\lambda u_{r})\dfrac{\partial Q}{\partial u_{t}}(t,r,u,\lambda u_{t},\lambda u_{r})\lambda du_{r} 
\end{equation*}
are differentiable with respect to $\lambda$ and vanish at $\lambda=0$.)

If a wave equation~\eqref{WErad} possesses a variational formulation~\eqref{variational}, 
then its multipliers $Q$ correspond to \emph{variational symmetries} defined by the characteristic form
\begin{equation}\label{varsymm}
 {\rm X}=r^{1-n}Q\partial/\partial u
\end{equation}
where the determining equations on $Q$ can be shown to reduce to conditions equivalent to those given by Noether's theorem
\cite{Olv,2ndbook,BluCheAnc}
for the Lagrangian functional ${\mathcal L}$ to be formally invariant 
(up to boundary terms at spatial infinity and the origin) 
under the prolongation of the generator \eqref{varsymm}. 
More precisely, 
symmetry invariance requires that the Lagrangian itself satisfies
\begin{equation}\label{invL}
 r^{n-1}\pr{\rm X}L=D_{t}A+D_{r}B
\end{equation}
for some scalar functions $A$ and $B$ depending on $t$, $r$, $u$, $u_{t}$, and $r$-derivatives of $u$, $u_{t}$,
where ${\rm X}$ is a symmetry generator \eqref{symgen} in characteristic form 
(and $\pr$ denotes its prolongation). 
Then the variational identity
\begin{equation}\label{noetherid}
 r^{n-1}\pr{\rm X}L=
\frac{\delta {\mathcal L}}{\delta u}P
+D_{t}\Big(r^{n-1}\frac{\partial L}{\partial u_{t}}P\Big)
+D_{r}\Big(r^{n-1}\frac{\partial L}{\partial u_{r}}P\Big) 
\end{equation}
yields the equation 
\begin{equation}\label{noetherthm}
\frac{\delta {\mathcal L}}{\delta u}P
= (u_{tt}-gu_{rr}-f)r^{n-1}P = D_{t}T +D_{r}X
\end{equation}
given by 
\begin{equation}\label{noetherconslaw}
T=A-Q\frac{\partial L}{\partial u_{t}},\quad 
X=B-Q\frac{\partial L}{\partial u_{r}},\quad
Q=r^{n-1}P.
\end{equation}
Thus, 
each symmetry generator \eqref{symgen} for which the Lagrangian functional is invariant \eqref{invL} 
yields a conservation law in characteristic form \eqref{characform}
with the conserved density $T$ and flux $X$ given in terms of $A,B,P$
by the Noether relation \eqref{noetherconslaw}.
Conversely, 
through the variational identity \eqref{noetherid} combined with equation \eqref{noetherthm}, 
each conservation law in characteristic form \eqref{characform} yields
a symmetry generator \eqref{varsymm} under which the Lagrangian functional is invariant \eqref{invL},
with $A$ and $B$ determined in terms of $Q,T,X$ 
from the Noether relation \eqref{noetherconslaw}.
Moreover, since invariance of the Lagrangian functional $\mathcal{L}$ implies that its stationary points $\delta \mathcal{L}/\delta u=0$
are preserved,
every variational symmetry \eqref{varsymm} of a Lagrangian for a wave equation \eqref{WErad}
is thus a symmetry of the wave equation itself.

Note that when the multiplier $Q$ given by a Noether conservation law \eqref{noetherconslaw} is of $1$st order \eqref{multiplierWE} 
then the corresponding variational symmetry \eqref{varsymm} will be of point type
if
\begin{equation}
 \dfrac{\partial^{2}Q}{\partial u_{t}\partial u_{t}}=\dfrac{\partial^{2}Q}{\partial u_{r}\partial u_{r}}
=\dfrac{\partial^{2}Q}{\partial u_{t}\partial u_{r}}=0
\end{equation}
and otherwise it will be of contact type.

By inspection, any Lagrangian functional of the form \eqref{variational} is formally invariant under
infinitesimal time-translation 
\begin{equation}\label{timetrans}
 \delta t=1,\ \delta r=0,\ \delta u=0.
\end{equation}
The corresponding characteristic generator thereby yields a variational symmetry
\begin{equation}
 {\rm X}=r^{1-n}Q\partial/\partial u,\quad Q=-r^{n-1}u_{t},
\end{equation}
admitted in the case of a variational wave equation \eqref{WErad}. The resulting conservation 
law \eqref{noetherconslaw} describes a conserved energy
\begin{equation}\label{energy}
 C=-\int_{0}^{\infty}T\; dr=\int_{0}^{\infty}(u_{t}^{2}+L(u,u_{t},u_{r}))r^{n-1}\; dr 
\end{equation}
for all solutions $u(t,r)$ that have zero flux at spatial infinity and at the origin. 

In the case of a wave equation \eqref{WEradc0} having the alternative variational structure \eqref{altvariational}, the previous
statements hold with $r^{m}$ in place of $r^{n-1}$.

For any wave equation \eqref{WErad}, whether or not it is variational, a conserved density $T$ and a flux $X$ that arise from a 
multiplier with at most linear dependence on $u_{t}$
\begin{equation}\label{genergyQ}
 Q=\alpha(t,r,u,u_{r})u_{t}+\beta(t,r,u,u_{r})
\end{equation}
will be called a \emph{generalized-energy} type conservation law. 
For such conservation laws, the relations \eqref{intmethodTX1}--\eqref{intmethodTX2} show that the form of $T$ and $X$ is given by
\begin{equation}\label{energyT}
\begin{aligned}
T=&
\tfrac{1}{2}u_{t}^{2}\alpha(t,r,u,u_{r})+u_{t}\beta(t,r,u,u_{r})+A(t,r,u,u_{r})\\
X=&
-\tfrac{1}{6}u_{t}^{3}\dfrac{\partial\alpha(t,r,u,u_{r})}{\partial u_{r}}
-\tfrac{1}{2}u_{t}^{2}\dfrac{\partial\beta(t,r,u,u_{r})}{\partial u_{r}}-u_{t}\dfrac{\partial A(t,r,u,u_{r})}{\partial u_{r}} +B(t,r,u,u_{r})
\end{aligned}
\end{equation}
for some scalar functions $A$ and $B$.

If a wave equation \eqref{WErad} possesses a divergence structure \eqref{divergence}, then it can be integrated to get
\begin{equation}
 \frac{d^{2}}{d t^{2}}\int_{0}^{\infty}ur^{n-1}\;dr=F\big|_{0}^{\infty}
\end{equation}
which formally yields a kinematic quantity given by
\begin{equation}\label{kinematicquantity}
 C(t)=\int_{0}^{\infty}ur^{n-1}\;dr=C_{1}t+C_{2}
\end{equation}
satisfying
\begin{equation}
 \frac{d^{2}C}{dt^{2}}=0
\end{equation}
provided that the flux $F$ vanishes at spatial infinity and at the origin.
The resulting quantities 
\begin{equation}\label{kinquantity1}
 C_{1}=\int_{0}^{\infty}u_{t}r^{n-1}\;dr=\text{const.}
\end{equation}
\begin{equation}\label{kinquantity2}
 C_{2}=\int_{0}^{\infty}(u-tu_{t})r^{n-1}\;dr=\text{const.}
\end{equation}
are conserved, 
corresponding to the conservation laws
\begin{equation}
 T_{1}=r^{n-1}u_{t},\quad X_{1}=-r^{n-1}F
\end{equation}
\begin{equation}
 T_{2}=r^{n-1}(u-tu_{t}),\quad X_{2}=r^{n-1}tF
\end{equation}
whose multipliers are given by
\begin{equation}
 Q_{1}=r^{n-1},\quad Q_{2}=-r^{n-1}t.
\end{equation}

In the case of a wave equation \eqref{WEradc0} having the alternative divergence structure \eqref{altdivergence}, 
the same kinematic conservation laws hold with $r^{m}$ in place of $r^{n-1}$.

More generally, 
for any wave equation \eqref{WErad} whether or not it has a divergence structure, 
a conserved density
$T$ and a flux $X$ arising from a multiplier $Q$ that depends only on the coordinates $t$, $r$ 
will be called a {\em kinematic} type conservation law. 
Such conservation laws are characterized by the form
\begin{equation}\label{kinematicT}
 T=u_{t}\alpha(t,r)-u\frac{\partial \alpha(t,r)}{\partial t}
\end{equation}
for the conserved density, where
\begin{equation}\label{kinematicQ}
 Q=\alpha(t,r)
\end{equation}
is the corresponding multiplier.

\section{Symmetry Classification}\label{sec:symmresults}

For each class of radial wave equations \eqref{WEreducrad1} and \eqref{WEreducrad2}
we will now find all point symmetries \eqref{varsymm}. 
In particular we will explicitly determine any point symmetries that exist only 
for special nonlinearity powers $p$ and dimensions $n\neq1$, 
as well as for special relations among the constant coefficients $a$, $b$, $c$ 
in these wave equations, 
excluding cases where the wave equation is linear or non-hyperbolic. 
(Note we will allow $n$ to have non-integer values. 
An interpretation of the wave equations \eqref{WEreducrad1} and \eqref{WEreducrad2} in such 
cases is given in section \ref{sec:concludingremarks}.)

Our results are obtained by solving the determining equation \eqref{detsymm} for the functions
$\tau(t,r,u)$, $\xi(t,r,u)$, $\eta(t,r,u)$ 
given by the characteristic form of the symmetry generator \eqref{pointsymm}.
To keep the list of solutions succinct, we have merged symmetries with similar forms into a combined form
wherever possible.
Remarks on the computation will be provided at the end 
in section~\ref{sec:symmcomputation}.

We will also work out the commutator structure of these symmetry generators. 
For this purpose it is convenient to use the associated canonical form of the generator
defined by 
\begin{equation}\label{canonicalX}
\hat{\rm X} = \tau\partial/\partial t + \xi\partial/\partial r + \eta\partial/\partial u
\end{equation}
which directly corresponds to an infinitesimal transformation \eqref{infinigen} on $(t,r,u)$, namely
$\delta t =\hat{\rm X} t =\tau(t,r,u)$, 
$\delta r =\hat{\rm X} r=\xi(t,r,u)$, 
$\delta u =\hat{\rm X} u=\eta(t,r,u)$.

\subsection{Point symmetries of $u_{tt}=(c+bu^{p})(u_{rr}+(n-1)u_{r}/r)+au^{p-1}u_{r}^{2}$.}\label{sec:Apointsymms}

\indent\newline\indent
Table~\ref{radAsym} lists the infinitesimal point symmetries of the radial wave equation \eqref{WEreducrad1} in all 
cases such that this equation is nonlinear (i.e.\ $a\neq0$ or $pb\neq0$), hyperbolic (i.e.\ $pb\neq0$ or $c+b\neq0$ when $pb=0$), 
and multi-dimensional (i.e.\ $n\neq1$). 
These three restrictions are equivalent to the inequality 
$(n-1)((bp)^2+a^2(c+b)^2)\neq0$ holding on the nonlinearity power $p$, the dimension $n$, and the coefficients $a$, $b$, $c$.

\begin{changemargin}{-2cm}{-1cm}
      \setcounter{tbn}{0}\setcounter{tabul}{0}
\begin{table}
      {\begin{center}\refstepcounter{tabul}\label{radAsym} 
      \setcounter{tbn}{0}
      \begin{tabular}{|c|c|c|c|c|}
      \hline\vspacebefore
      \hfill  &\hfill $\tau\hfill$ &\hfill $\xi\hfill$ &\hfill$\eta\hfill$ &\hfill {conditions\hfill} \\
      \hline\vspacebeforemore
      $1$ 
	& $1$ 
	& $0$ 
	& $0$ 
	&  - \\[1.2ex]
      \hline\vspacebeforemore
      $2$ 
	& $t$ 
	& $r$ 
	& $0$ 
	&  -  \\[1.2ex]
      \hline\vspacebeforemore
      $3$ 
	& $0$ 
	& $\tfrac{1}{2}pbr$ 
	& $bu+pc$  
	& $cp(p-1)=0,\ b\neq0$ \\[1.2ex]
      \hline\vspacebeforemore      
      $4$ 
	& $0$ 
	& $\big(a+b+(1-p)c\big)r^{3-n}$ 
	& $\begin{aligned}(2-n)r^{2-n}\big((b+(1-p)c)u+pc\big)\end{aligned}$  
	& $\begin{aligned}&cp(p-1)=0,\\& \big(a+b+(1-p)c\big)(n-3)\\&\quad=bp(n/2-1),\\& n\neq2\end{aligned}$ \\[1.2ex]
      \hline\vspacebeforemore
      $5$ 
	& $0$ 
	& $\tfrac{1}{2}pbr\ln{r}$ 
	& $(1+\ln{r})(bu+pc)$  
	& $\begin{aligned}&cp(p-1)=0,\\ &a=(p-1)c-b, b\neq0,\\ &n=2\end{aligned}$ \\[1.2ex]
      \hline\vspacebeforemore
      $6$ 
	& $t^{2}$ 
	& $0$ 
	& $tu$  
	& $\begin{aligned}&c=0,\\& p=-4\end{aligned}$ \\[1.2ex]
      \hline\vspacebeforemore
      $7$ 
	& $0$ 
	& $0$ 
	& $1$ 
	& $\begin{aligned}&b=0,\\& p=1\end{aligned}$  \\[1.2ex]
      \hline\vspacebeforemore
      $8$ 
	& $0$ 
	& $0$ 
	& $t$ 
	& $\begin{aligned}&b=0,\\& p=1\end{aligned}$  \\[1.2ex]
      \hline
      \end{tabular}
      \end{center}}
\caption{Infinitesimal point symmetries for $n\neq1$, $(bp)^2+a^2(c+b)^2\neq0$.}
\end{table}
\end{changemargin}

Symmetries $\hatrmX{1}$ and $\hatrmX{2}$ respectively generate a time-translation and a space-time dilation on $(t,r)$. 
These are the only symmetries admitted for all allowed values of $a$, $b$, $c$, $p$, $n$. 
Symmetries $\hatrmX{7}$ and $\hatrmX{8}$ generate a shift on $u$, while symmetry $\hatrmX{6}$ generates a temporal inversion on $(t,u)$. 
Symmetry $\hatrmX{3}$ generates a scaling on $(r,u)$ when $c=0$, or a scaling on $u$ when $p=0$, 
and otherwise generates a shift on $u$ combined with a scaling on $(r,u)$ when $c\neq0$, $p\neq0$. 
Symmetry $\hatrmX{4}$ is a non-rigid generalization of the shift-scaling $\hatrmX{3}$ on $(r,u)$.
Symmetry $\hatrmX{5}$ is a logarithmic counterpart of $\hatrmX{4}$. 

From Table~\ref{radAsym} we see that the point symmetry structure for the wave equation \eqref{WEreducrad1} is richest in four main cases: 
$n=2$, i.e.\ when the spatial domain is planar; $p=1$, i.e.\ when the nonlinearities are quadratic;
$p=0$, i.e. when the Laplacian term is linear;
and $c=0$, i.e.\ when the linear Laplacian term is absent. 
We also see that the only distinguished relation among $a$, $b$, $c$ is given by
$(a+b+(1-p)c)(n-3)=bp(n/2-1)$, 
which reduces to $a=(p-1)c-b$ when $n=2$, where $p$ and $c$
are restricted by the condition $p(p-1)c=0$. 
When the radial wave equation \eqref{WEreducrad1} is quasilinear (i.e.\ $p\neq0$, $b\neq0$), this relation becomes
\begin{equation}\label{quasiabrelation}
a=-\Big(p\dfrac{1-n/2}{n-3}+1\Big)b,
\quad n\neq3
\end{equation}
with $p=1$ or $c=0$.
In contrast, when the radial wave equation \eqref{WEreducrad1} is semilinear (i.e.\ 
$pb=0$, $a\neq0$), the distinguished relation among $a$, $b$, $c$ simplifies to
\begin{equation}\label{semiabrelation}
a=-(b+c) \text{ or }  n=3
\end{equation}
We note that, from Proposition~\ref{structures}, 
the simplified relation \eqref{quasiabrelation} in the quasilinear case coincides with 
the condition $2a=pb$ for existence of a variational structure \eqref{variational} if $p=2(n-3) \neq0$
or the condition $a=pb$ for existence of a divergence structure \eqref{divergence} if $p=2(n-3)/(4-n)\neq 0$, $n\neq4$,
while the non-existence of both a variational structure and a divergence structure
in the semilinear case precludes any overlap with the relation \eqref{semiabrelation}. 

We will next summarize the structure of the algebra generated by all of these point symmetries.

\begin{thm}\label{Aquasilinsymmalgebra}
For a multi-dimensional, hyperbolic, quasilinear radial wave equation \eqref{WEreducrad1}, 
its point symmetries in the general case $n\neq1$, $p\neq0$, $b\neq0$ comprise a 2-dimensional algebra generated by 
the time-translation $\hatrmX{1}$ and the space-time dilation $\hatrmX{2}$, 
with the commutator structure 
\begin{equation}\label{Aquasilincom1}
[\hatrmX{1},\hatrmX{2}]=\hatrmX{1}
\end{equation}
Its additional point symmetries generate larger algebras in the following cases:
\newline
{\rm (i)}
For $c=0$, the point symmetries additionally comprise 
the scaling $\hatrmX{3}$;
the non-rigid dilation $\hatrmX{4}$ when 
$a(n-3)+b(n-3 -p(n-2)/2)=0$, $n\neq2$; 
the logarithmic dilation $\hatrmX{5}$ when $a+b=0$, $n=2$; 
and the temporal inversion $\hatrmX{6}$ when $p=-4$. 
In the subcases 
$n=3$, $p\neq 0$, $p\neq -4$, 
or $n\neq3$, $a\neq -b\big(p(1-n/2)/(n-3)+1\big)$, $p\neq -4$, 
the scaling $\hatrmX{3}$ enlarges the commutator structure \eqref{Aquasilincom1} by
\begin{equation}\label{Aquasilincom2}
[\hatrmX{1},\hatrmX{3}]=[\hatrmX{2},\hatrmX{3}]=0
\end{equation}
generating a 3-dimensional algebra. 
In the subcases 
$n\neq2$, $n\neq3$, $a=-b\big(p(1-n/2)/(n-3)+1\big)$, $p\neq-4$, 
or $n=3$, $p=0$, 
the non-rigid dilation $\hatrmX{4}$ enlarges the commutator structure \eqref{Aquasilincom1}--\eqref{Aquasilincom2} by 
\begin{equation}\label{Aquasilincom3}
[\hatrmX{1},\hatrmX{4}]=0,
[\hatrmX{2},\hatrmX{4}]=(2-n)\hatrmX{4},
[\hatrmX{3},\hatrmX{4}]=p(1-n/2)b\hatrmX{4}.
\end{equation}
In the subcase 
$n=2$, $a=-b$, $p\neq-4$, 
the logarithmic dilation $\hatrmX{5}$ enlarges the commutator structure \eqref{Aquasilincom1}--\eqref{Aquasilincom2} by
\begin{equation}\label{Aquasilincom4}
[\hatrmX{1},\hatrmX{5}]=0,
[\hatrmX{2},\hatrmX{5}]=\hatrmX{3},
[\hatrmX{3},\hatrmX{5}]=pb/2\hatrmX{3}.
\end{equation}
In the subcase 
$n\neq2$, $n\neq3$, $a\neq -b\big(p(1-n/2)/(n-3)+1\big)$, $p=-4$, 
the temporal inversion $\hatrmX{6}$ enlarges the commutator structure \eqref{Aquasilincom1}--\eqref{Aquasilincom2} by 
\begin{equation}\label{Aquasilincom5}
[\hatrmX{1},\hatrmX{6}]=2\hatrmX{2}+1/b\hatrmX{3},
[\hatrmX{2},\hatrmX{6}]=\hatrmX{6},
[\hatrmX{3},\hatrmX{6}]=0.
\end{equation}
In the three previous subcases, the algebra is 4-dimensional. 
A 5-dimensional algebra is generated in the two remaining subcases. 
In the subcase 
$n\neq2$, $n\neq3$, $a=-b\big((4-2n)/(n-3)+1\big)$, $p=-4$, 
the non-rigid dilation $\hatrmX{4}$ and the temporal inversion $\hatrmX{6}$ enlarge the commutator structure \eqref{Aquasilincom1}--\eqref{Aquasilincom2} by 
\begin{equation}\label{Aquasilincom6}
[\hatrmX{1},\hatrmX{4}]=[\hatrmX{4},\hatrmX{6}]=0,
[\hatrmX{2},\hatrmX{4}]=(2-n)\hatrmX{4},
[\hatrmX{3},\hatrmX{4}]=2(n-2)b\hatrmX{4}.
\end{equation}
In the subcase $n=2$, $a=-b$, $p=-4$, 
the logarithmic dilation $\hatrmX{5}$ and the temporal inversion $\hatrmX{6}$ enlarge the commutator structure \eqref{Aquasilincom1}--\eqref{Aquasilincom2} by 
\begin{equation}\label{Aquasilincom7}
[\hatrmX{1},\hatrmX{5}]=[\hatrmX{5},\hatrmX{6}]=0,
[\hatrmX{2},\hatrmX{5}]=\hatrmX{3},
[\hatrmX{3},\hatrmX{5}]=-2b\hatrmX{3}.
\end{equation}
{\rm (ii)}
For $c\neq0$, $p=1$, 
the point symmetries additionally comprise the shift-scaling $\hatrmX{3}$;
the non-rigid dilation $\hatrmX{4}$ when $a=-b\big((1-n/2)/(n-3)+1\big)$, $n\neq2$, $n\neq3$; 
and the logarithmic dilation $\hatrmX{5}$ when $a=-b$, $n=2$. 
In the subcase 
$n\neq3$, $a\neq -b\big((1-n/2)/(n-3)+1\big)$,
the shift-scaling $\hatrmX{3}$ enlarges the commutator structure to generate a 3-dimensional algebra \eqref{Aquasilincom1}--\eqref{Aquasilincom2}. 
A 4-dimensional algebra is generated in the two remaining subcases. 
In the subcase $n\neq2$, $n\neq3$, $a=-b\big((1-n/2)/(n-3)+1\big)$,  
the non-rigid dilation $\hatrmX{4}$ enlarges the commutator structure \eqref{Aquasilincom1}--\eqref{Aquasilincom2} by 
\begin{equation}\label{Aquasilincom8}
[\hatrmX{1},\hatrmX{4}]=0,
[\hatrmX{2},\hatrmX{4}]=(2-n)\hatrmX{4},
[\hatrmX{3},\hatrmX{4}]=(1-n/2)b\hatrmX{4}.
\end{equation}
In the subcase 
$n=2$, $a=-b$, 
the logarithmic dilation $\hatrmX{5}$ enlarges the commutator structure 
\eqref{Aquasilincom1}--\eqref{Aquasilincom2} by
\begin{equation}\label{Aquasilincom9}
[\hatrmX{1},\hatrmX{5}]=0,
[\hatrmX{2},\hatrmX{5}]=\hatrmX{3},
[\hatrmX{3},\hatrmX{5}]=b/2\hatrmX{3}.
\end{equation}
\end{thm}

Variational point symmetries of a radial wave equation \eqref{WEreducrad1} comprise a
sub-algebra of the algebra described in Theorem~\ref{Aquasilinsymmalgebra}. 
The structure of this sub-algebra will be discussed in the context of conservation laws
in section~\ref{sec:Aconslaws}. 

\begin{thm}\label{Asemilinsymmalgebra}
For a multi-dimensional, hyperbolic, semilinear radial wave equation \eqref{WEreducrad1}, 
its point symmetries in the general case $n\neq1$, $bp=0$, $a\neq0$, $c+b\neq0$ again comprise a 2-dimensional algebra generated by 
the time-translation $\hatrmX{1}$ and the space-time dilation $\hatrmX{2}$, 
with the commutator structure 
\begin{equation}\label{Asemilincom1}
[\hatrmX{1},\hatrmX{2}]=\hatrmX{1}.
\end{equation}
Its additional point symmetries generate larger algebras in following cases: 
\newline
{\rm (i)}
For $p=0$, $b\neq0$, 
the point symmetries additionally comprise 
the scaling $\hatrmX{3}$; 
the non-rigid dilation $\hatrmX{4}$ when $(a+b+c)(n-3)=0$, $n\neq2$; 
and the logarithmic dilation $\hatrmX{5}$ when $a+b+c=0$, $n=2$. 
In the subcase 
$n\neq3$, $a\neq-(b+c)$, 
the scaling $\hatrmX{3}$ enlarges the commutator structure \eqref{Asemilincom1} by
\begin{equation}\label{Asemilincom2}
[\hatrmX{1},\hatrmX{3}]=[\hatrmX{2},\hatrmX{3}]=0
\end{equation}
generating a 3-dimensional algebra. 
In the subcases 
$n=3$, or $n\neq2$, $n\neq3$, $a=-(b+c)$, 
the scaling $\hatrmX{3}$ and the non-rigid dilation $\hatrmX{4}$ enlarge the commutator structure \eqref{Asemilincom1} and \eqref{Asemilincom2} by
\begin{equation}\label{Asemilincom3}
[\hatrmX{1},\hatrmX{4}]=[\hatrmX{3},\hatrmX{4}]=0,
[\hatrmX{2},\hatrmX{4}]=(2-n)\hatrmX{4}
\end{equation}
generating a 4-dimensional algebra. 
In the subcase 
$n=2$, $a=-(b+c)$, 
the scaling $\hatrmX{3}$ and the logarithmic dilation $\hatrmX{5}$ enlarge the commutator structure \eqref{Asemilincom1} and \eqref{Asemilincom2} by 
\begin{equation}\label{Asemilincom4}
[\hatrmX{1},\hatrmX{5}]=[\hatrmX{3},\hatrmX{5}]=0,
[\hatrmX{2},\hatrmX{5}]=\hatrmX{3}
\end{equation}
generating another 4-dimensional algebra.
\newline
{\rm (ii)}
For $b=0$, $c\neq0$, 
the point symmetries additionally comprise 
the non-rigid dilation $\hatrmX{4}$ when $(a+(1-p)c)(n-3)=0$, $p(p-1)=0$, $n\neq2$;
and the shifts $\hatrmX{7}$ and $\hatrmX{8}$ when $p=1$. 
In the subcases
$p=0$, $n=3$, or $p=0$, $n\neq3$, $a=-c$, 
the non-rigid dilation $\hatrmX{4}$ enlarges the commutator structure \eqref{Asemilincom1} by 
\begin{equation}\label{Asemilincom5}
[\hatrmX{1},\hatrmX{4}]=0,
[\hatrmX{2},\hatrmX{4}]=(2-n)\hatrmX{4}
\end{equation}
generating a 3-dimensional algebra. 
In the subcase $p=1$, $n\neq3$, the shifts $\hatrmX{7}$ and $\hatrmX{8}$ enlarge the commutator structure \eqref{Asemilincom1} by 
\begin{equation}\label{Asemilincom6}
[\hatrmX{1},\hatrmX{7}]=[\hatrmX{2},\hatrmX{7}]=0,
[\hatrmX{1},\hatrmX{8}]=\hatrmX{7},
[\hatrmX{2},\hatrmX{8}]=\hatrmX{8}
\end{equation}
generating a 4-dimensional algebra. 
In the subcase $p=1$, $n=3$, 
the non-rigid dilation $\hatrmX{4}$ and the shifts $\hatrmX{7}$ and $\hatrmX{8}$ 
enlarge the commutator structure \eqref{Asemilincom1} and \eqref{Asemilincom6} by 
\begin{equation}\label{Asemilincom7}
[\hatrmX{1},\hatrmX{4}]=[\hatrmX{4},\hatrmX{7}]=[\hatrmX{4},\hatrmX{8}]=0,
[\hatrmX{2},\hatrmX{4}]=-\hatrmX{4}
\end{equation}
generating a 5-dimensional algebra.
\end{thm}

Since semilinearity for a radial wave equation \eqref{WEreducrad1} 
precludes the existence of a variational formulation, 
the algebra described in Theorem~\ref{Asemilinsymmalgebra} has no variational sub-algebra.

\subsection{Point symmetries of $u_{tt}=(c+(a+b)u_{r}^{p})u_{rr}+(n-1)(cu_{r}+bu_{r}^{p+1})/r$.}\label{sec:Bpointsymms}

\indent\newline\indent
Table~\ref{radBsymm} lists the infinitesimal point symmetries of the radial wave equation \eqref{WEreducrad2} in all cases 
such that this equation is nonlinear (i.e.\ $pb\neq0$ or $p(a+b)\neq0$), 
hyperbolic (i.e.\ $p(b+a)\neq0$ or $c\neq0$), 
and multi-dimensional (i.e.\ $n\neq1$, $c\neq0$ or $(p+1)b\neq0$). 
These three restrictions are equivalent to the inequality 
$(n-1)p(b^2+c^2)((b+a)^2+b^2c^2(p+1)^2)\neq0$ holding on the nonlinearity power $p$, the dimension $n$, and the coefficients $a$, $b$, $c$.

\begin{table}
\begin{center}\refstepcounter{tabul}\label{radBsymm}
      \setcounter{tbn}{0}
      \begin{tabular}{|c|c|c|c|c|c|}
      \hline\vspacebefore
      \hfill  &\hfill $\tau\hfill$ &\hfill $\xi\hfill$ &\hfill$\eta\hfill$ &\hfill {conditions\hfill} \\
      \hline\vspacebeforemore
      $1$ 
	& $1$ 
	& $0$ 
	& $0$ 
	&  -  \\[1.2ex]
      \hline\vspacebeforemore
      $2$ 
	& $t$ 
	& $r$ 
	& $u$ 
	&  -  \\[1.2ex]
      \hline\vspacebeforemore
      $3$ 
	& $0$ 
	& $0$ 
	& $1$  
	&  - \\[1.2ex]
      \hline\vspacebeforemore
      $4$ 
	& $0$ 
	& $0$ 
	& $t$  
	&  - \\[1.2ex]
      \hline\vspacebeforemore
      $5$ 
	& $pt$ 
	& $0$ 
	& $-2u$  
	& $c=0$ \\[1.2ex]
      \hline\vspacebeforemore
      $6$ 
	& $t^{2}$ 
	& $0$ 
	& $tu$  
	& $\begin{aligned}&c=0,\\& p=-4\end{aligned}$ \\[1.2ex]
      \hline\vspacebeforemore
      $7$ 
	& $0$ 
	& $r^{-m}$ 
	& $0$  
	& $\begin{aligned}&c=0,\ a\neq b(n-2),\ b\neq0 \\& p=-2\end{aligned}$ \\[1.2ex]
      \hline\vspacebeforemore
      $8$ 
	& $0$ 
	& $r\ln{r}$ 
	& $0$  
	& $\begin{aligned}&c=0,\ a=b(n-2),\ b\neq0,\\&  p=-2\end{aligned}$ \\[1.2ex]
      \hline
      \end{tabular}
      \end{center}
\caption{Infinitesimal point symmetries for $(n-1)(b^2(p+1)^2+c^2)\neq0$, $p\big((b+a)^2+b^2c^2(p+1)^2\big)\neq0$.}
\end{table}
Symmetry $\hatrmX{1}$ generates a time-translation. 
Symmetry $\hatrmX{2}$ generates a scaling on $(t,r,u)$. 
Symmetries $\hatrmX{3}$ and $\hatrmX{4}$ generate a shift on $u$.
These are the only symmetries admitted for all allowed values of $a$, $b$, $c$, $p$, $n$. 
Symmetry $\hatrmX{5}$ generates a scaling on $(t,u)$, while symmetry $\hatrmX{6}$ generates a temporal inversion on $(t,u)$. 
Symmetry $\hatrmX{7}$ is a non-rigid generalization of the scaling on $r$
given by $\hatrmX{2}+\tfrac{1}{2}\hatrmX{5}$ (i.e.\ $\tau=0$, $\xi=r$, $\eta=0$) when $c=0$.
Symmetry $\hatrmX{8}$ is a logarithmic counterpart of $\hatrmX{7}$. 

From table~\ref{radBsymm} we see that the point symmetry structure for the wave equation \eqref{WEreducrad2} is richest 
in the case $c=0$, i.e.\ when the linear Laplacian term is absent, 
and in the subcases $p=-2$ and $p=-4$, i.e.\ when the nonlinearities take the form
\begin{equation}
 u_{tt}=\dfrac{a+b}{|p|}(u_{r}^{-|p|})_{r}+(n-1)bu_{r}^{-|p+1|}/r,\quad 
p=-2,4.
\end{equation}
We also see that there is no distinguished dimension $n\neq1$,
while the only distinguished relation among $a$, $b$, $c$ is given by
\begin{equation}\label{Bspecrelation}
 a=b(n-2),\ b\neq0.
\end{equation}
When the radial wave equation \eqref{WEreducrad2} is quasilinear (i.e.\ $p\neq0$, $b+a\neq0$), this relation \eqref{Bspecrelation} 
coincides with the condition $a=pb$ for existence of both 
a variational structure \eqref{variational} and a divergence structure 
\eqref{divergence} if $p=n-2\neq0$. 
Moreover, the relation \eqref{Bspecrelation} corresponds to
\begin{equation}
 m=p+1 
\end{equation}
in terms of the radial power 
\begin{equation}
 m=\dfrac{b(n-1)(p+1)}{a+b}
\end{equation}
which characterizes the alternative variational and divergence structures
\eqref{altvariational} and \eqref{altdivergence} that exist for
the radial wave equation \eqref{WEreducrad2} in the highly quasilinear case $c=0$ ($p\neq0$, $b+a\neq0$). 
In contrast, when the radial wave equation \eqref{WEreducrad2} is semilinear (i.e.\ 
$p\neq0$, $b+a=0$, $bc\neq0$), 
the non-existence of both a variational structure and a divergence structure 
precludes any overlap with the relation \eqref{Bspecrelation}. 

We will next summarize the structure of the algebra generated by all of these point symmetries.

\begin{thm}\label{Bquasilinsymmalgebra}
For a multi-dimensional, hyperbolic, quasilinear radial wave equation \eqref{WEreducrad2}, 
its point symmetries in the general case $n\neq1$, $b\neq0$, $b+a\neq0$, $p(p+1)\neq0$ 
comprise a 4-dimensional algebra generated by 
the time-translation $\hatrmX{1}$, 
the scaling $\hatrmX{2}$, 
and the shifts $\hatrmX{3}$ and $\hatrmX{4}$, 
with the commutator structure 
\begin{equation}\label{Bquasilincom1}
[\hatrmX{1},\hatrmX{2}]=\hatrmX{1},
[\hatrmX{1},\hatrmX{3}]=[\hatrmX{2},\hatrmX{4}]=[\hatrmX{3},\hatrmX{4}]=0,
[\hatrmX{2},\hatrmX{3}]=-\hatrmX{3}.
\end{equation}
For $c=0$, 
its point symmetries generate larger algebras that additionally comprise 
the temporal scaling $\hatrmX{5}$; 
the temporal inversion $\hatrmX{6}$ when $p=-4$;
the non-rigid scaling $\hatrmX{7}$ when $a\neq b(n-2)$, $p=-2$;
and the logarithmic scaling $\hatrmX{8}$ when $a=b(n-2)$, $p=-2$. 
In the subcase 
$p\neq-2$, $p\neq-4$, 
the temporal scaling $\hatrmX{5}$ enlarges the commutator structure \eqref{Bquasilincom1} by
\begin{equation}\label{Bquasilincom2}
[\hatrmX{1},\hatrmX{5}]=p\hatrmX{1},
[\hatrmX{2},\hatrmX{5}]=0,
[\hatrmX{3},\hatrmX{5}]=-2\hatrmX{3}, 
[\hatrmX{4},\hatrmX{5}]=-(2+p)\hatrmX{4}
\end{equation}
generating a 5-dimensional algebra. 
In the subcase $b\neq0$, $p=-4$, 
the temporal inversion $\hatrmX{6}$ enlarges the 
commutator structure \eqref{Bquasilincom1} and \eqref{Bquasilincom2} by
\begin{equation}\label{Bquasilincom3}
[\hatrmX{1},\hatrmX{6}]=-\tfrac{1}{2}\hatrmX{5},
[\hatrmX{2},\hatrmX{6}]=\hatrmX{6},
[\hatrmX{3},\hatrmX{6}]=\hatrmX{4},
[\hatrmX{4},\hatrmX{6}]=0,
[\hatrmX{5},\hatrmX{6}]=-4\hatrmX{6}.
\end{equation}
In the subcase $a\neq b(n-2)$, $p=-2$, 
the non-rigid scaling $\hatrmX{7}$ enlarges the commutator structure
\eqref{Bquasilincom1} and \eqref{Bquasilincom2} by
\begin{equation}\label{Bquasilincom5}
[\hatrmX{1},\hatrmX{7}]=[\hatrmX{3},\hatrmX{7}]=[\hatrmX{4},\hatrmX{7}]=[\hatrmX{5},\hatrmX{7}]=0,
[\hatrmX{2},\hatrmX{7}]=-(m+1)\hatrmX{7}.
\end{equation}
In the subcase $a=b(n-2)$, $p=-2$, 
the logarithmic scaling $\hatrmX{8}$ enlarges the commutator structure
\eqref{Bquasilincom1} and \eqref{Bquasilincom2} by
\begin{equation}\label{Bquasilincom6}
[\hatrmX{1},\hatrmX{8}]=[\hatrmX{3},\hatrmX{8}]=[\hatrmX{4},\hatrmX{8}]=[\hatrmX{5},\hatrmX{8}]=0,
[\hatrmX{2},\hatrmX{8}]=\hatrmX{2}+\tfrac{1}{2}\hatrmX{5}.
\end{equation}
In the previous three subcases, the algebra is 6-dimensional. 
\end{thm}

Variational point symmetries of a radial wave equation \eqref{WEreducrad2} comprise a
sub-algebra of the algebra described in Theorem~\ref{Bquasilinsymmalgebra}. 
The structure of this sub-algebra will be discussed in the context of conservation laws
in section~\ref{sec:Bconslaws}. 

\begin{thm}\label{Bsemilinsymmalgebra}
For a multi-dimensional, hyperbolic, semilinear radial wave equation \eqref{WEreducrad2}, 
its point symmetries in all cases $n\neq1$, $b+a=0$, $b\neq0$, $c\neq0$, $p(p+1)\neq0$ 
comprise a 4-dimensional algebra generated by 
the time-translation $\hatrmX{1}$, 
the scaling $\hatrmX{2}$, 
and the shifts $\hatrmX{3}$ and $\hatrmX{4}$, 
with the commutator structure 
\begin{equation}\label{Bsemilincom1}
[\hatrmX{1},\hatrmX{2}]=\hatrmX{1},
[\hatrmX{1},\hatrmX{3}]=[\hatrmX{2},\hatrmX{4}]=[\hatrmX{3},\hatrmX{4}]=0,
[\hatrmX{1},\hatrmX{3}]=\hatrmX{3},
[\hatrmX{2},\hatrmX{3}]=-\hatrmX{3}.
\end{equation}
\end{thm}

Since the existence of a variational formulation is precluded by semilinearity 
for a radial wave equation \eqref{WEreducrad2}, 
the algebra described in Theorem~\ref{Bsemilinsymmalgebra} has no variational sub-algebra.

\subsection{Computational Remarks.}\label{sec:symmcomputation}

\indent\newline\indent
In the computation of point symmetries \eqref{pointsymm} for a radial wave equation \eqref{WEreducrad}, 
the symmetry determining equation \eqref{detsymm} is formulated in the jet space 
whose coordinates are defined by $t$, $r$, $u$, $u_{t}$, $u_{r}$, $u_{tr}$ and $u_{rr}$, 
where $u_{tt}$ is replaced by $gu_{rr}+f$.
For wave equation \eqref{WEreducrad1}, 
this determining equation splits with respect to $u_{t}$, $u_{r}$, $u_{tr}$, $u_{rr}$ 
into an overdetermined system of 16 equations in 8 unknowns $\eta(t,r,u)$, $\xi(t,r,u)$, $\tau(t,r,u)$, $a$, $b$, $c$, $n$, and $p$.
In contrast, for wave equation \eqref{WEreducrad2}, 
the determining equation splits initially with respect to 
$u_{t}$, $u_{tr}$ and $u_{rr}$, 
but not with respect to $u_{r}$ 
because this wave equation contains terms involving the unknown power $p$ of $u_{r}$.
In this case the partial splitting yields an overdetermined system of 9 equations in the 8 unknowns 
$\eta(t,r,u)$, $\xi(t,r,u)$, $\tau(t,r,u)$, $a$, $b$, $c$, $n$, and $p$.

We use computer algebra both to carry out the splitting and to solve the resulting overdetermined system of equations.
The formulation and splitting of the symmetry determining equation is done automatically 
by the program {\sc LiePDE} \cite{Wolf1993} 
which in turn calls the package {\sc Crack} \cite{Wolf2002a} 
to solve the resulting overdetermined system for the 8 unknowns.
Although the system is linear in the unknowns $\eta$, $\tau$ and $\xi$,
it is nonlinear jointly in the 8 unknowns and even is non-polynomial in the unknown power $p$ which appears as an exponent.
Such problems are typically much more complicated to solve than are linear problems.

One main difficulty is the handling of unknowns as exponents, 
which requires case splittings such that all possible balances of exponents are considered. 
An example is the equation 
$\xi_{t}-(a+b)u_{r}^{p}\tau_{r}-c\tau_{r}-(a+b)u_{r}^{p+1}\tau_{u}-cu_{r}\tau_{u}=0$ 
together with the inequality 
$p\neq0$ 
arising during the computation for solving the overdetermined system in the case of wave equation \eqref{WEreducrad2}.
Balancing the powers $0$, $1$, $p$, $p+1$ of $u_{r}$ leads to three different splittings:
$p=-1$; $p=1$; $(p-1)(p+1)\neq0$.
Each of these three cases then must be solved separately in the remaining computation.
A similar situation arises during the computation for solving the overdetermined system for wave equation \eqref{WEreducrad1}, 
which involves splitting with respect to powers of $u$ after integrating out the $u$ dependence of $\tau$, $\xi$, $\eta$.

Other kinds of case splittings also arise. 
For example, when an expression factorizes then each factor must be considered zero or non-zero, 
as in the equation $c(\xi_{r}-\tau_{t})=0$ 
arising during the computation for solving the overdetermined system for wave equation \eqref{WEreducrad2}. 
This leads to the cases $c=0$ and  $c\neq0$, $\xi_{r}=\tau_{t}$, 
both of which must then be solved separately.

To our knowledge, 
{\sc Crack} is the only package that can handle such computational difficulties 
and, in particular, can both generate and solve all resulting cases automatically.
In handling nested case distinctions, 
{\sc Crack} considers equations and inequalities together. 
For example, equations and inequalities are simplified with each other, 
and new inequalities as well as equations are generated through case distinctions.

For both wave equations \eqref{WEreducrad1} and \eqref{WEreducrad2}, 
{\sc Crack} is able to complete the entire computation of point symmetries automatically.

\section{Conservation law Classification}\label{sec:conslawresults}

For each class of radial wave equations \eqref{WEreducrad1} and \eqref{WEreducrad2}
we will now find all kinematic conservation laws \eqref{kinematicT}--\eqref{kinematicQ} 
and all generalized-energy conservation laws \eqref{genergyQ}--\eqref{energyT}. 
In particular we will explicitly determine any such conservation laws that exist only 
for special nonlinearity powers $p$ and dimensions $n\neq1$, 
as well as for special relations among the constant coefficients $a$, $b$, $c$ 
in these wave equations, excluding cases where the wave equation is linear or non-hyperbolic.
(Note we will allow $n$ to have non-integer values. 
An interpretation of the wave equations \eqref{WEreducrad1} and \eqref{WEreducrad2} in such 
cases is given in section \ref{sec:concludingremarks}.)

Our results are obtained by first solving the determining system \eqref{detcon1}--\eqref{detcon2} for the multiplier function 
\begin{equation}\label{classifyQ}
Q=\alpha(t,r,u,u_{r})u_{t}+\beta(t,r,u,u_{r})
\end{equation}
given by the characteristic form \eqref{characform} of these conservation laws, 
and then using the homotopy integration formula \eqref{TfromQ}--\eqref{WforTX} or directly integrating relations
\eqref{intmethodTX1}--\eqref{intmethodTX2}
to obtain the conserved density $T(t,r,u,u_{t},u_{r})$ and flux $X(t,r,u,u_{t},u_{r})$ 
determined by each multiplier.
To keep the list of solutions succinct, we have merged conservation laws with similar forms into a combined form
wherever possible.
Remarks on the computation will be provided at the end
in section~\ref{sec:conslawcomputation}.

In the case when a radial wave equation \eqref{WEreducrad1} or \eqref{WEreducrad2} possesses a variational formulation \eqref{variational}, 
multipliers \eqref{classifyQ} correspond 
(via Noether's theorem \eqref{noetherid}--\eqref{noetherthm}) 
to symmetries with the characteristic form
\begin{equation}\label{classifyX}
{\rm X}=r^{1-n}(\alpha(t,r,u,u_{r})u_{t}+\beta(t,r,u,u_{r}))\partial/\partial u.
\end{equation}
Thus, our classification of all generalized-energy conservation laws will provide
a partial classification of variational symmetries for these wave equations.
We note that a symmetry generator \eqref{classifyX} will be of point type \eqref{pointsymm} iff
\begin{equation}\label{pointconds}
\frac{\partial\alpha}{\partial u_r}=\frac{\partial^2\beta}{\partial u_r\partial u_r}=0
\end{equation}
in which case it will have the equivalent canonical form 
\begin{equation}
\hat{\rm X} = 
-r^{1-n}\Big(\alpha\partial/\partial t 
+\frac{\partial\beta}{\partial u_r}\partial/\partial r 
+\big(\frac{\partial\beta}{\partial u_r} u_r-\beta\big)\partial/\partial u\Big).
\end{equation}
In the case of the alternative variational structure \eqref{altvariational}--\eqref{m},
the same statements hold with $r^{1-n}$ replaced by $r^{-m}$.

\subsection{Conservation laws of $u_{tt}=(c+bu^{p})(u_{rr}+(n-1)u_{r}/r)+au^{p-1}u_{r}^{2}$.}\label{sec:Aconslaws}

\indent\newline\indent
In the following tables we list the kinematic conservation laws \eqref{kinematicT}--\eqref{kinematicQ}
and generalized-energy conservation laws \eqref{genergyQ}--\eqref{energyT}
of the radial wave equation \eqref{WEreducrad1} 
in all cases such that this equation is
nonlinear (i.e.\ $a\neq0$ or $pb\neq0$), 
hyperbolic (i.e.\ $pb\neq0$ or $c+b\neq0$ when $pb=0$), 
and multi-dimensional (i.e.\ $n\neq1$). 
These three restrictions are equivalent to the inequality 
$(n-1)((bp)^2+a^2(c+b)^2)\neq0$ holding on the nonlinearity power $p$, the dimension $n$, and the coefficients $a$, $b$, $c$.

To begin, 
the kinematic conserved densities, fluxes, and corresponding multipliers 
are listed in Table~\ref{radAconslaw0th}, 
where $H(u)$ denotes the expression \eqref{intcasesplit} appearing in the divergence structure \eqref{flux1}.

\begin{table}
\begin{center}\refstepcounter{tabul}\label{radAconslaw0th}
      \setcounter{tbn}{0}
      \begin{tabular}{|c|c|c|c|c|}
      \hline\vspacebefore
      \hfill &\hfill $Q\hfill$ &\hfill $T\hfill$ & \hfill $X\hfill$ &\hfill {conditions\hfill} \\
      \hline\vspacebeforemore
      $1$ 
	& $r^{n-1}$ 
	& $r^{n-1}u_{t}$ 
	& $-r^{n-1}(c+bu^{p})u_{r} $ 
	& $a=pb$ \\[1.2ex]
      \hline\vspacebeforemore
      $2$ 
	& $r^{n-1}t$ 
	& $r^{n-1}(tu_{t}-u)$ 
	& $-r^{n-1}t(c+bu^{p})u_{r} $ 
	& $a=pb$ \\[1.2ex]
      \hline\vspacebeforemore
      $3$ 
	& $r$ 
	& $ru_{t}$ 
	& $-r(c+bu^{p})u_{r}-(n-2)(cu+bH(u))$ 
	& $\begin{aligned}&a=pb,\\& n\neq2\end{aligned}$ \\[1.2ex]
      \hline\vspacebeforemore
      $4$ 
	& $rt$ 
	& $r(tu_{t}-u)$ 
	& $-t\big(r(c+bu^{p})u_{r}+(n-2)(cu+bH(u))\big) $ 
	& $\begin{aligned}&a=pb,\\ & n\neq2\end{aligned}$ \\[1.2ex]
      \hline\vspacebeforemore
      $5$ 
	& $r\ln r$ 
	& $(r\ln r)u_{t}$ 
	& $cu+bH(u)-r(\ln{r})(c+bu^{p})u_{r} $ 
	& $\begin{aligned}&a=pb,\\& n=2\end{aligned}$ \\[1.2ex]
      \hline\vspacebeforemore
      $6$ 
	& $(r\ln r)t$
	& $(r\ln r)(tu_{t}-u)$ 
	& $t\big(cu+bH(u)-r(\ln{r})(c+bu^{p})u_{r}\big) $ 
	& $\begin{aligned}&a=pb,\\& n=2\end{aligned}$ \\[1.2ex]
      \hline
      \end{tabular}
      \end{center}
 \caption{Kinematic conservation laws for $n\neq1$, $(bp)^2+a^2(c+b)^2\neq0$.}
\end{table}
Conserved densities $T_1$ and $T_2$ 
arise from the $n$-dimensional divergence structure \eqref{divergence} 
for the radial wave equation \eqref{WEreducrad1}. 
Conserved densities $T_3$ and $T_4$ 
arise from an alternative divergence structure 
$(ru_t)_t = (rF)_r$ 
where the flux is given by $F=(c+bu^p) u_r +(n-2)(cu+bH(u))$,
which holds under the same existence condition $a=pb$ 
as for the divergence structure \eqref{divergence}. 
A logarithmic counterpart of the alternative divergence structure gives rise to the conserved densities $T_5$ and $T_6$.

We will now summarize the structure of the vector space spanned by the conserved quantities \eqref{C} arising from these kinematic conserved densities. 

\begin{thm}\label{Akinconslawsvs}
A multi-dimensional, hyperbolic, nonlinear radial wave equation \eqref{WEreducrad1}
admits kinematic conservation laws only in cases when $a=pb\neq0$. 
\newline
{\rm (i)}
For $n\neq2$, $a=pb\neq0$, 
a basis of kinematic conserved quantities is comprised by
\begin{equation}\label{Akinbasis1}
C_1=\int_0^\infty u_{t}r^{n-1} dr, \quad
C_2=\int_0^\infty (tu_{t}-u)r^{n-1} dr, \quad
C_3=\int_0^\infty u_{t}r dr, \quad
C_4=\int_0^\infty (tu_{t}-u)r dr.
\end{equation}
\newline
{\rm (ii)}
For $n=2$, $a=pb\neq0$, 
a basis of kinematic conserved quantities is comprised by 
\begin{equation}\label{Akinbasis2}
C_1=\int_0^\infty u_{t}r dr, \quad
C_2=\int_0^\infty (tu_{t}-u)r dr,\quad
C_5=\int_0^\infty u_{t} (\ln r)r dr,\quad
C_6=\int_0^\infty (tu_{t}-u)(\ln r)r dr.
\end{equation}
\newline
In both cases the conserved quantities span a 4-dimensional vector space. 
\end{thm}

The conditions on the coefficients $a,b,c$ coming from Theorem~\ref{Akinconslawsvs}
combined with Proposition~\ref{structures}
imply the following interesting result. 

\begin{cor}
A multi-dimensional, hyperbolic, nonlinear radial wave equation \eqref{WEreducrad1}
possesses kinematic conservation laws iff it has a quasilinear divergence structure. 
\end{cor}

Next, the generalized-energy conserved densities and fluxes are listed in Table~\ref{radAconslaw1st};
corresponding multipliers are listed in Table~\ref{radAmultiplier1st}.

\begin{table}
\begin{changemargin}{-2cm}{-1cm}
 {\begin{center}\refstepcounter{tabul}\label{radAconslaw1st}
      \setcounter{tbn}{0}
      \begin{tabular}{|c|c|c|c|c|}
      \hline\vspacebefore
      \hfill &\hfill $T\hfill$ &\hfill $X\hfill$ &\hfill {conditions\hfill} \\
      \hline\vspacebeforemore
      $1$ 
	& $\tfrac{1}{2}r^{n-1}(u_{t}^{2}+(c+bu^{p})u_{r}^{2})$
	& $-r^{n-1}(c+bu^{p})u_{r}u_{t} $
        & $2a=bp$ \\[1.2ex]
      \hline\vspacebeforemore
      $2$ 
	& $\begin{aligned}\tfrac{1}{2}r^{n-1}\big(
               &b(np+4)t(u_{t}^{2} +(c+bu^p)u_{r}^2)\\
               &+2b(p+4)ru_{r}u_{t}\\&+4(n-1)(c+bu)u_{t}\big) 
            \end{aligned}$ 
        & $\begin{aligned}-\tfrac{1}{2}r^{n-1}\big(&b(p+4)r((c+bu)u_{r}^{2}+u_{t}^{2})\\
               &+2(c+bu^{p})(b(np+4)tu_{t}\\&+2(n-1)(c+bu))u_{r}\big)
             \end{aligned}$
        & $2a=bp$, $c(p-1)=0$ \\[1.2ex]
      \hline\vspacebeforemore
      $3$ 
	& $r(bru_{r}+bu+c)u_{t} $ 
        & $\begin{aligned}-\tfrac{1}{2}\big(&(c+bu^{p})(ru_{r}+u)(bru_{r}+bu+2c)\\&+br^2u_{t}^2+c^2u\big)
           \end{aligned}$
        & $\begin{aligned}&2a=bp,\ c(p-1)=0,\\& p=2(n-3)\end{aligned}$ \\[1.2ex]
      \hline\vspacebeforemore
      $4$ 
	& $\tfrac{1}{2}r^{n-1}\big(t^{2}(u_t^2+bu^{-4}u_r^2)
             -2tuu_t +u^2\big)$ 
        & $-r^{n-1}btu^{-4}u_{r}(tu_{t}-u) $
        &  $\begin{aligned}&2a=bp,\ c=0,\\& p=-4\end{aligned}$\\[1.2ex]
      \hline\vspacebeforemore
      $5$ 
	& $\tfrac{1}{2}rt(u_{t}^{2}+bu^{-2}u_{r}^{2}) +(r\ln{r})(ru_{r}+u)u_{t}$ 
        & $\begin{aligned}-\tfrac{1}{2}\big(&(r\ln{r})(r(u_{t}^{2}+bu^{-2}u_{r}^{2})+2bu^{-1}u_{r})\\
              &+2b(rtu^{-2}u_{t}u_{r}-\ln{u})\big)
             \end{aligned}$
        & $\begin{aligned}&2a=bp,\ c=0,\\& p=-2,\ n=2\end{aligned}$ \\[1.2ex]
      \hline\vspacebeforemore
      $6$ 
	& $\begin{aligned}\tfrac{1}{2}r^{n-1}\big(&(3bru_{r}+n(bu+c))u_{t}^{2}\\
              &+(bu^p+c)(bru_{r}+n(bu+c))u_{r}^{2}\big)\end{aligned}$ 
        & $\begin{aligned}-\tfrac{1}{2}r^{n-1}u_{t}\big(&2n(c+bu)(c+bu^{p})u_{r}\\
              &+r(3b(c+bu^{p})u_{r}^{2}+bu_{t}^{2})\big)
             \end{aligned}$
        &  $\begin{aligned}&3a=bp,\ c(p-1)=0,\\& p=6(1-3/n)\end{aligned}$ \\[1.2ex]
      \hline\vspacebeforemore
      $7$ 
	& $\begin{aligned}\tfrac{1}{2}r^{4/5}\big(
              &(5ru_r+3u)(tu_t^2-2uu_t)\\&+3btu^{-4}(5u_{r}+9u)u_{r}^{2}\big)
             \end{aligned}$ 
        & $\begin{aligned}-\tfrac{1}{2}r^{4/5}\big(&u_{r}(5bru_{r}+6bu)(tu_{t}-u)u^{-4}\\
              &+\tfrac{5}{3}ru_{t}^{2}(tu_{t}-3u)\big)
             \end{aligned}$
        &  $\begin{aligned}&3a=bp,\ c=0,\\& p=-4,\ n=9/5\end{aligned}$ \\[1.2ex]
      \hline
      \end{tabular}
      \end{center}}
\caption{Generalized-energy conservation laws for $n\neq1$, $(bp)^2+a^2(c+b)^2\neq0$.}
\end{changemargin}
\end{table}

\begin{table}
\begin{changemargin}{-2cm}{-1cm}
 {\begin{center}\refstepcounter{tabul}\label{radAmultiplier1st}
      \setcounter{tbn}{0}
      \begin{tabular}{|c|c|c|c|c|}
      \hline\vspacebefore
      \hfill &\hfill $Q\hfill$ &\hfill {conditions\hfill} \\
      \hline\vspacebeforemore
      $1$ 
	& $r^{n-1}u_{t}$ 
	& $2a=bp$ \\[1.2ex]
      \hline\vspacebeforemore
      $2$ 
	& $\begin{aligned} 
              r^{n-1}\big(&b(np+4)tu_{t}+b(p+4)ru_{r}\\ &+2(n-1)(c+bu)\big) 
            \end{aligned}$ 
	& $2a=bp$, $c(p-1)=0$ \\[1.2ex]
      \hline\vspacebeforemore
      $3$ 
	&$r(bru_{r}+bu+c)$ 
	& $\begin{aligned}&2a=bp,\ c(p-1)=0,\\& p=2(n-3)\end{aligned}$ \\[1.2ex]
      \hline\vspacebeforemore
      $4$ 
	&$r^{n-1}t(tu_t-u)$ 
        &  $\begin{aligned}&2a=bp,\ c=0,\\& p=-4\end{aligned}$\\[1.2ex]
      \hline\vspacebeforemore
      $5$ 
	&$rtu_{t}+(r\ln{r})(ru_r+u)$ 
        & $\begin{aligned}&2a=bp,\ c=0,\\& p=-2,\ n=2\end{aligned}$ \\[1.2ex]
      \hline\vspacebeforemore
      $6$ 
	&  $r^{n-1}u_t(3bru_r+n(bu+c))$ 
        &  $\begin{aligned}&3a=bp,\ c(p-1)=0,\\& p=6(1-3/n)\end{aligned}$ \\[1.2ex]
      \hline\vspacebeforemore
      $7$ 
	&  $r^{4/5}(tu_t-u)(5ru_r+3u)$ 
        &  $\begin{aligned}&3a=bp,\ c=0,\\& p=-4,\ n=9/5\end{aligned}$ \\[1.2ex]
      \hline
      \end{tabular}
      \end{center}}
\caption{Generalized-energy multipliers for $n\neq1$, $(bp)^2+a^2(c+b)^2\neq0$.}
\end{changemargin}
\end{table}

From Table~\ref{radAconslaw1st} we see that these conservation laws are characterized by two distinguished relations among 
the coefficients $a,b,c$: $2a=pb$, or $3a=pb$. 
The first relation $2a=pb$ is simply the condition for existence of 
a variational structure \eqref{variational} for the radial wave equation \eqref{WEreducrad1}.
Conservation laws in this case correspond to variational symmetries through Noether's theorem,
as we will summarize in the next subsection~\ref{sec:ANoether}.  
The second relation $3a=pb$ turns out to be the condition needed for deriving 
a variational type of Morawetz radial dilation identity 
which produces conservation laws from temporal symmetries. 
We will discuss this interesting case later in subsection~\ref{sec:AMorawetz}.  

\subsubsection{Noether correspondence for 
$u_{tt}=(c+bu^{p})(u_{rr}+(n-1)u_{r}/r)+(bp/2)u^{p-1}u_{r}^{2}$.}\label{sec:ANoether}  

In the variational case $2a=pb$, 
multipliers $Q$ for conservation laws of the radial wave equation \eqref{WEreducrad1}
are in one-to-one correspondence to variational symmetries ${\rm X}=r^{1-n}Q\partial/\partial u$ 
of the Lagrangian \eqref{Lagrang1}. 
We see from Table~\ref{radAmultiplier1st} that 
all of the generalized-energy multipliers for $2a=pb$ are linear in $u_t$ and $u_r$. 
Consequently, the corresponding variational symmetries are all of point type. 
This correspondence is summarized in Table~\ref{radAgenerators}, 
where $C$ denotes the conserved quantity \eqref{C} given by the conservation laws, 
and $\hat{\rm X}$ denotes the variational symmetry generator in canonical form \eqref{canonicalX}. 

\begin{table}
\begin{changemargin}{-2cm}{-1cm}
 {\begin{center}\refstepcounter{tabul}\label{radAgenerators}
      \setcounter{tbn}{0}
      \begin{tabular}{|c|c|c|c|c|}
      \hline\vspacebefore
      \hfill &\hfill $\hat{\rm X}\hfill$ &\hfill $C\hfill$  &\hfill {conditions\hfill} \\
      \hline\vspacebeforemore
      $1$ 
	& $-\partial/\partial t$
	& $\displaystyle\int_{0}^{\infty}\tfrac{1}{2}(u_{t}^{2}+(c+bu^{p})u_{r}^{2})r^{n-1}\;dr$
        & - \\[1.2ex]
      \hline\vspacebeforemore
      $2$ 
	& $\begin{aligned} 
              &-b(np+4)t\partial/\partial t-b(p+4)r\partial/\partial r\\ &+2(n-1)(c+bu)\partial/\partial u 
            \end{aligned}$ 
	& $\begin{aligned}\displaystyle\int_{0}^{\infty}\tfrac{1}{2}\big(
               &b(np+4)t(u_{t}^{2} +(c+bu^p)u_{r}^2)\\
               &+2b(p+4)ru_{r}u_{t}\\&+4(n-1)(c+bu)u_{t}\big) 
               r^{n-1}\;dr\end{aligned}$
        & $c(p-1)=0$ \\[1.2ex]
      \hline\vspacebeforemore
      $3$ 
	& $-br^{3-n}\partial/\partial r+r^{2-n}(bu+c)\partial/\partial u$ 
	& $\displaystyle\int_{0}^{\infty}(bru_{r}+bu+c)u_{t}r\;dr$
	& $\begin{aligned}&c(p-1)=0,\\& p=2(n-3)\end{aligned}$ \\[1.2ex]
      \hline\vspacebeforemore
      $4$ 
	& $-t^2\partial/\partial t-tu\partial/\partial u$ 
	& $\begin{aligned}\displaystyle\int_{0}^{\infty}\tfrac{1}{2}\big(&t^{2}(u_t^2+bu^{-4}u_r^2)\\
               &-2tuu_t +u^2\big)r^{n-1}\;dr\end{aligned}$
        & $\begin{aligned}&c=0,\\& p=-4\end{aligned}$\\[1.2ex]
      \hline\vspacebeforemore
      $5$ 
	& $-t\partial/\partial t+r\ln{r}\partial/\partial r+ (\ln{r})u\partial/\partial u$ 
	& $\begin{aligned}\displaystyle\int_{0}^{\infty}\big(\tfrac{1}{2}t(&u_{t}^{2}+bu^{-2}u_{r}^{2})\\  
	      &+(\ln{r})(ru_{r}+u)u_{t}\big)r\;dr
	     \end{aligned}$
        & $\begin{aligned}&c=0,\\& p=-2,\ n=2\end{aligned}$ \\[1.2ex]
      \hline
      \end{tabular}
      \end{center}}
\caption{Generalized-energy conserved quantities and variational symmetries for $n\neq1$, $2a=pb\neq0$.}
\end{changemargin}
\end{table}

Conserved quantity $C_1$ is an energy arising from the time-translation symmetry $\hatrmX{1}$. 
Conserved quantity $C_2$ is a similarity energy arising from the scaling symmetry $\hatrmX{2}$, 
while conserved quantity $C_5$ is a logarithmic counterpart of $C_2$. 
Conserved quantity $C_3$ is a radial momentum arising from the non-rigid radial dilation symmetry $\hatrmX{3}$. 
Conserved quantity $C_4$ is a conformal energy arising from the temporal inversion symmetry $\hatrmX{4}$. 

We will now state the vector space structure spanned by 
these conserved quantities as well as the algebra structure 
determined by the corresponding variational symmetries. 

\begin{thm}\label{Anoetherconslawsvs}
For a multi-dimensional, hyperbolic, nonlinear radial wave equation \eqref{WEreducrad1}
that has a variational structure \eqref{variational} when $2a=pb\neq0$, 
variational point symmetries are admitted in two cases, 
corresponding to all generalized-energy conservation laws. 
\newline
{\rm (i)}
For $c=0$, the variational point symmetries comprise 
the time-translation $\hatrmX{1}$; the scaling $\hatrmX{2}$;
the non-rigid radial dilations $\hatrmX{3}$ when $p=2(n-3)$ 
and $\hatrmX{5}$ when $p=-2$, $n=2$;
and the temporal inversion $\hatrmX{4}$ when $p=-4$.
In the subcase $p\neq 2(n-3)$, $p\neq-4$, 
the time-translation $\hatrmX{1}$ and the scaling $\hatrmX{2}$ generate 
a two-dimensional algebra with the commutator structure 
\begin{equation}\label{Anoethercom1}
[\hatrmX{1},\hatrmX{2}]=-b(np+4)\hatrmX{1}.
\end{equation}
The corresponding two-dimensional vector space of conserved quantities is spanned by 
the energy $C_1$ and the similarity energy $C_2$. 
In the subcase $p=2(n-3)$, $p\neq-4$, $n\neq2$, 
the non-rigid radial dilation $\hatrmX{3}$ enlarges the commutator structure \eqref{Anoethercom1} by 
\begin{equation}\label{Anoethercom2}
[\hatrmX{1},\hatrmX{3}]=0,
[\hatrmX{2},\hatrmX{3}]=2b(n-1)(n-2)\hatrmX{3}
\end{equation}
generating a three-dimensional algebra, 
while the corresponding vector space of conserved quantities is spanned by 
$C_1$, $C_2$, and the radial momentum $C_3$. 
In the subcase $p=-4$,
the temporal inversion $\hatrmX{4}$ enlarges the commutator structure \eqref{Anoethercom1} by 
\begin{equation}\label{Anoethercom3}
[\hatrmX{1},\hatrmX{4}]=-1/(4b(n-1))\hatrmX{2},
[\hatrmX{2},\hatrmX{4}]=4b(n-1)\hatrmX{4}
\end{equation}
generating another three-dimensional algebra, 
while the corresponding vector space of conserved quantities is spanned by 
$C_1$, $C_2$, and the conformal energy $C_4$. 
In the subcase $p=-2$, $n=2$, 
the non-rigid radial dilations $\hatrmX{3}$ and $\hatrmX{5}$ 
enlarge the commutator structure \eqref{Anoethercom1}--\eqref{Anoethercom2} by 
\begin{equation}\label{Anoethercom4}
[\hatrmX{1},\hatrmX{5}]=-\hatrmX{1},
[\hatrmX{2},\hatrmX{5}]=2\hatrmX{3},
[\hatrmX{3},\hatrmX{5}]=\hatrmX{3}
\end{equation}
generating a four-dimensional algebra whose corresponding vector space of conserved quantities is spanned by 
$C_1$, $C_2$, $C_3$, and the dilational energy $C_5$. 
\newline
{\rm (ii)}
For $c\neq0$, $p=1$, the variational point symmetries comprise 
the time-translation $\hatrmX{1}$; the scaling $\hatrmX{2}$;
and the non-rigid radial dilation $\hatrmX{3}$ when $n=7/2$. 
In the subcase $n\neq7/2$, 
the time-translation $\hatrmX{1}$ and the scaling $\hatrmX{2}$ generate 
a two-dimensional algebra with the commutator structure \eqref{Anoethercom1},
while the corresponding vector space of conserved quantities is spanned by 
the energy $C_1$ and the similarity energy $C_2$. 
In the subcase $n=7/2$, 
the non-rigid radial dilation $\hatrmX{3}$ enlarges the commutator structure \eqref{Anoethercom1} by 
\begin{equation}\label{Anoethercom5}
[\hatrmX{1},\hatrmX{3}]=0,
[\hatrmX{2},\hatrmX{3}]=15/2b\hatrmX{3}
\end{equation}
generating a three-dimensional algebra, 
while the corresponding vector space of conserved quantities is spanned by 
$C_1$, $C_2$, and the radial momentum $C_3$. 
\end{thm}

\subsubsection{Morawetz-type variational identity for 
$u_{tt}=(c+bu^{p})(u_{rr}+(n-1)u_{r}/r)+(bp/3)u^{p-1}u_{r}^{2}$.}\label{sec:AMorawetz}

To derive a variational identity for the radial wave equation \eqref{WEreducrad1} when $2a\neq pb$, 
we start by making use of the previous Lagrangian \eqref{Lagrang1} to write 
\begin{equation}\label{WELrad1}
r^{n-1}(u_{tt}-(c+bu^p)(u_{rr}+(n-1)u_{r}/r)-a u^{p-1}u_{r}^{2})
= E_u(r^{n-1} L) +\mu r^{n-1}\frac{\partial L}{\partial u}
\end{equation}
where 
\begin{equation}\label{nonvarcoeff1}
\mu= 1-\frac{2a}{pb}\neq0
\end{equation}
is the coefficient of the non-variational terms, 
and where 
\begin{equation}\label{EuLoperator}
E_u = \partial/\partial u -D_r\partial/\partial u_r -D_t\partial/\partial u_t
\end{equation}
is the 1st order Euler-Lagrange operator (i.e.\ local variational derivative)
with respect to $u$. 
We next multiply equation \eqref{WELrad1} by the expression 
$W=ru_r-\lambda u$ which represents the characteristic function 
given by the generator of a scaling transformation on $(r,u)$, namely 
$\delta t=0$, $\delta r=r$, $\delta u=\lambda u$, 
for some constant $\lambda$. 
This yields
\begin{align}\label{PEL1}
& 
r^{n-1}(ru_r-\lambda u)\big(u_{tt}-(c+bu^p)(u_{rr}+(n-1)u_{r}/r)-a u^{p-1}u_{r}^{2}\big)
\nonumber\\&
= (ru_r-\lambda u)E_u(r^{n-1} L) +\mu r^{n-1}(ru_r-\lambda u)\frac{\partial L}{\partial u}.
\end{align}
Now we will show that the Euler-Lagrange term in this equation can be expressed
in a useful way as the sum of a total variational derivative $E_u(r^{n-1} \tilde L)$
for some new Lagrangian $\tilde L(u,u_r,u_t)$ 
and a dilational derivative $(rD_r+\tilde\lambda)(r^{n-1} f)$
for some function $f(u,u_t)$ and some constant $\tilde\lambda$.
Our main tool is the operator identity 
\begin{equation}\label{operid}
WE_u = \pr{\rm Y} -D_t(W\partial/\partial u_t) -D_r(W\partial/\partial u_r) 
\end{equation}
where $\pr{\rm Y}$ is the prolongation of the generator ${\rm Y}=W\partial/\partial u$.

To proceed we split up the Lagrangian $L=L_0+L_1$ into its temporal and radial parts
\begin{equation}\label{splitL1} 
L_0= -\tfrac{1}{2}u_{t}^2 ,\quad
L_1= \tfrac{1}{2}(c+bu^{p})u_{r}^2
\end{equation}
which have the scaling properties
\begin{equation}\label{scalL0term1}
\pr{\rm Y}L_0= rD_r L_0 -2\lambda L_0 
\end{equation}
and 
\begin{equation}\label{scalL1term1}
\pr{\rm Y}L_1= 
rD_r L_1 -2(\lambda-1)L_1 -\lambda u\frac{\partial L_1}{\partial u}
\end{equation}
for $W=ru_r-\lambda u$. 
Applying the identity \eqref{operid} to the part of the Euler-Lagrange term containing $L_0$ in equation \eqref{PEL1}, 
and using the scaling property \eqref{scalL0term1}, 
we get 
\begin{equation}\label{simpL0term1}
(ru_r-\lambda u)E_u(r^{n-1} L_0)
= D_t\big(r^{n-1}(ru_r-\lambda u)u_t\big) 
+r^{n-1}(rD_r-2\lambda)L_0.
\end{equation}
If we now define ${\tilde L}_0$ by 
\begin{equation}
\frac{\partial {\tilde L}_0}{\partial u_t} = (ru_r-\lambda u)u_t
\end{equation}
then the time derivative term in equation \eqref{simpL0term1}
can be expressed as 
\begin{equation}\label{Dtterm1}
D_t\big(r^{n-1}(ru_r-\lambda u)u_t\big) 
= -E_u(r^{n-1} {\tilde L}_0) +r^{n-1}(rD_r+(\lambda+n))L_0
\end{equation}
where
\begin{equation}\label{L0new1}
{\tilde L}_0 = \tfrac{1}{2}(ru_r-\lambda u)u_{t}^{2} .
\end{equation}
Hence equation \eqref{simpL0term1} becomes
\begin{equation}\label{L0term1}
(ru_r-\lambda u)E_u(r^{n-1} L_0)
= -E_u(r^{n-1} {\tilde L}_0) +r^{n-1}(2rD_r+(n-\lambda))L_0.
\end{equation}
Similarly, from the scaling property \eqref{scalL1term1} combined with 
the identity \eqref{operid} applied to the part of the Euler-Lagrange term containing $L_1$ in equation \eqref{PEL1},
we obtain 
\begin{equation}\label{simpL1term1}
(ru_r-\lambda u)E_u(r^{n-1} L_1)
= -D_r\big(r^{n-1}(ru_r-\lambda u)(c+bu^{p})u_r\big)
+r^{n-1}\big( (rD_r-2(\lambda-1))L_1 -\lambda u\frac{\partial L_1}{\partial u} \big) .
\end{equation}
If we integrate by parts on the dilational derivative term in equation \eqref{simpL1term1}
and then define ${\tilde L}_1$ by 
\begin{equation}
\frac{\partial {\tilde L}_1}{\partial u_r} 
= -\tfrac{1}{2}(ru_r-2\lambda u)(c+bu^{p})u_r , 
\end{equation}
we find that the radial derivative terms in equation \eqref{simpL1term1}
can be expressed as 
\begin{equation}\label{Drterm1}
-D_r\big(r^{n-1}(ru_r-\lambda u)(c+bu^{p})u_r\big) +r^{n}D_r L_1
= -E_u(r^{n-1} {\tilde L}_1) +r^{n-1}\big( (\lambda(p+1)-n)L_1 
-\tfrac{1}{3}ru_r\frac{\partial L_1}{\partial u} \big) 
\end{equation}
where 
\begin{equation}\label{L1new1}
{\tilde L}_1 = -\tfrac{1}{6}(ru_r-3\lambda u)(c+bu^{p})u_r^2.
\end{equation}
Hence equation \eqref{simpL1term1} becomes
\begin{equation}\label{L1term1}
(ru_r-\lambda u)E_u(r^{n-1} L_1)
= -E_u(r^{n-1} {\tilde L}_1) +r^{n-1}\big( (2-n-\lambda)L_1 
-\tfrac{1}{3}ru_r \frac{\partial L_1}{\partial u} \big) .
\end{equation}
Finally, substituting expressions \eqref{L0term1} and \eqref{L1term1} into 
equation \eqref{PEL1}, we have
\begin{align}\label{simpPEL1}
& 
r^{n-1}(ru_r-\lambda u)\big(u_{tt}-(c+bu^p)(u_{rr}+(n-1)u_{r}/r)-a u^{p-1}u_{r}^{2}\big)
\nonumber\\&
= -E_u(r^{n-1}({\tilde L}_0 + {\tilde L}_1)) +r^{n-1}(2rD_r +(n-\lambda))L_0
\nonumber\\&\qquad
+r^{n-1}\big( (2-n-\lambda)L_1 
+( (\mu-\tfrac{1}{3})ru_r -\mu\lambda u) \frac{\partial L_1}{\partial u} \big) .
\end{align}
We now require that the terms $L_1$, $u_r\partial L_1/\partial u$, $u\partial L_1/\partial u$ vanish, which holds iff the coefficients of 
$u^{p-1}u_{r}^3$, $u_{r}^2$, $u^{p}u_{r}^2$ in this equation are equal to zero. 
For $p\neq 0$, this imposes the respective conditions
\begin{equation}\label{conds1}
\mu=1/3,\quad
c(2-n-\lambda)=0 ,\quad
(1+p/3)\lambda=2-n.
\end{equation}
From the expression \eqref{nonvarcoeff1} for $\mu$ in terms of $a,b$, we then obtain 
\begin{equation}\label{coeff1}
3a=pb.
\end{equation}

Thus, under conditions \eqref{conds1}--\eqref{coeff1},
we see that equation \eqref{PEL1} reduces to the variational dilation identity
\begin{align}\label{morawetz1}
& 
r^{n-1}(ru_r-\lambda u)\big(u_{tt}-(c+bu^p)(u_{rr}+(n-1)u_{r}/r)-(pb/3) u^{p-1}u_{r}^{2}\big)
\nonumber\\&
= -E_u(r^{n-1}{\tilde L}) -(rD_r+\tilde\lambda)(r^{n-1}u_t^2)
\end{align}
with 
\begin{equation}\label{Lnew1}
{\tilde L} = {\tilde L}_0 + {\tilde L}_1 
= \tfrac{1}{2}(ru_r-\lambda u)u_{t}^{2} -\tfrac{1}{6}(ru_r-3\lambda u)(c+bu^{p})u_r^2
\end{equation}
and 
\begin{equation}\label{constnew1}
\tilde\lambda= 1-(n+\lambda)/2.
\end{equation}
As we will now illustrate, 
this identity \eqref{morawetz1} is able to produce conservation laws 
from multipliers that are given by certain symmetries 
${\rm X} = P\partial/\partial u$ in characteristic form, 
in a manner similar to Noether's theorem \eqref{noetherid}--\eqref{noetherthm}. 

Consider the time translation 
\begin{equation}
{\rm X} = u_t\partial/\partial u.
\end{equation}
When the identity \eqref{morawetz1} is multiplied by 
\begin{equation}\label{symmchar} 
 P = u_t
\end{equation} 
we see first that 
the Euler-Lagrange term yields a total time derivative
\begin{equation}
-u_t E_u(r^{n-1}{\tilde L}) 
= D_t\big( r^{n-1}u_t \frac{\partial \tilde L}{\partial u}\big)
- r^{n-1}\pr{\rm X} \tilde L
= D_t\big( r^{n-1}u_t \frac{\partial \tilde L}{\partial u}-r^{n-1}\tilde L \big)
\end{equation}
since, by inspection, 
the Lagrangian functional 
$\tilde{\mathcal L}=\int_{t_0}^{t_1}\int_0^\infty {\tilde L}r^{n-1} drdt$ 
is invariant under time-translation. 
Second, we see that the dilational derivative term produces 
\begin{equation}
-u_t (rD_r +\tilde\lambda)(r^{n-1}u_t^2)
= -2r^n u_t^2 u_{tr} -(\tilde\lambda +n -1) r^{n-1} u_t^3
= D_r\big( -\tfrac{2}{3} r^n u_t^3 \big) + (1-\tfrac{1}{3} n -\tilde\lambda)r^{n-1} u_t^3
\end{equation}
which will reduce to a total radial derivative iff 
\begin{equation}\label{constnewcond1}
\tilde\lambda= 1-n/3.
\end{equation}
Equating expressions \eqref{constnewcond1} and \eqref{constnew1}, 
and using conditions \eqref{conds1}, 
we obtain the additional conditions
\begin{equation}\label{pcond1}
c=0,\quad 
p=6(1-3/n) 
\end{equation}
as well as 
\begin{equation}\label{scalcond1}
\lambda= -n/3.
\end{equation}

Thus, for $a$, $b$, $c$, $p$, $\lambda$ satisfying conditions \eqref{coeff1}, \eqref{constnewcond1} and \eqref{pcond1},
the product of the variational dilation identity \eqref{morawetz1} and the symmetry characteristic function 
\eqref{symmchar} yields a conservation law, 
where the conserved density and flux are given by $T_6$ and $X_6$ in Table~\ref{radAconslaw1st}
while the multiplier is given by $Q_6$ in Table~\ref{radAmultiplier1st}. 
We note that the form of this multiplier consists of $r^{n-1}$ times the product of 
the characteristic functions $W=ru_r-\lambda u$ and $P = u_t$ 
representing the scaling transformation and the time-translation symmetry respectively.
The $c\neq 0$ case for this conservation law 
shown in Tables~\ref{radAconslaw1st} and~\ref{radAmultiplier1st}
can be obtained from a similar variational identity in which the scaling transformation
is generalized to a shift-scaling transformation 
given by a characteristic function of the form $W=ru_r-\lambda (u+\nu)$,
for some constants $\lambda\neq0$ and $\nu\neq0$. 
In addition, the remaining conservation law given by 
the conserved density $T_7$ and flux $X_7$ in Table~\ref{radAconslaw1st}, 
with the multiplier given by $Q_7$ in Table~\ref{radAmultiplier1st}, 
can be obtained from the variational identity \eqref{morawetz1} by using 
the temporal scaling symmetry with the characteristic function $P = tu_t-u$
in place of the time-translation symmetry. 
These results are summarized in Table~\ref{radAgeneratorsMorowitz}, 
where $C$ denotes the conserved quantity \eqref{C} given by the conservation laws, 
and where $\hat{\rm Y}$ and $\hat{\rm X}$ respectively denote 
the scaling generator and symmetry generator in canonical form \eqref{canonicalX}. 

\begin{table}
\begin{changemargin}{-2cm}{-1cm}
 {\begin{center}\refstepcounter{tabul}\label{radAgeneratorsMorowitz}
      \setcounter{tbn}{0}
      \begin{tabular}{|c|c|c|c|c|}
      \hline\vspacebefore
      \hfill &\hfill $\hat{\rm Y}\hfill$ &\hfill $\hat{\rm X}\hfill$ &\hfill $C\hfill$  &\hfill {conditions\hfill} \\
      \hline\vspacebeforemore
      $6$ 
	&  $\begin{aligned}&n(bu+c)\partial/\partial u\\
                &-3br\partial/\partial r
            \end{aligned}$
        &  $-\partial/\partial t$ 
        &  $\begin{aligned}\displaystyle\int_{0}^{\infty}\tfrac{1}{2}\big(&(3bru_{r}+n(bu+c))u_{t}^{2}\\
            &+(bu^p+c)(bru_{r}+n(bu+c))u_{r}^{2}\big)r^{n-1}\;dr
           \end{aligned}$
        &  $\begin{aligned}&c(p-1)=0,\\& p=6(1-3/n)\end{aligned}$ \\[1.2ex]
      \hline\vspacebeforemore
      $7$ 
	&  $3u\partial/\partial u-5r\partial/\partial r$
        &  $-u\partial/\partial u-t\partial/\partial t$
        & $\begin{aligned}\displaystyle\int_{0}^{\infty}\tfrac{1}{2}\big(&u_{r}(5bru_{r}+6bu)(u-tu_{t})u^{-4}\\
            &+\tfrac{5}{3}ru_{t}^{2}(3u-tu_{t})\big)r^{4/5}\;dr
           \end{aligned}$
        &  $\begin{aligned}&c=0,\\& p=-4,\ n=9/5\end{aligned}$ \\[1.2ex]
      \hline
      \end{tabular}
      \end{center}}
\caption{Generalized-energy conserved quantities and generators for $n\neq1$, $3a=bp\neq0$.}
\end{changemargin}
\end{table}

We will now state the vector space structure spanned by 
these conserved quantities. 

\begin{thm}\label{Amorawetzconslawsvs}
For a multi-dimensional, hyperbolic, nonlinear radial wave equation \eqref{WEreducrad1}
that has no variational structure \eqref{variational}, 
generalized-energy conservation laws are admitted only in cases when $3a=pb\neq0$, 
arising from a variational type of radial dilation identity. 
\newline
{\rm (i)}
For $c=0$, $3a=pb\neq0$, the conserved quantities are comprised by 
the generalized-dilation energy $C_6$ when $p=6(1-3/n)$ 
and the generalized-dilation conformal energy $C_7$ when $p=-4$, $n=9/5$. 
In the subcase $p=6(1-3/n)$, $n\neq 9/5$, 
a basis for the conserved quantities consists of $C_6$, 
spanning a 1-dimensional vector space. 
In the subcase $p=-4$, $n=9/5$, 
the basis consists of $C_6$, $C_7$, 
spanning a 2-dimensional vector space. 
\newline
{\rm (ii)}
For $c\neq0$, $3a=pb\neq0$, $p=1$, the conserved quantities comprise only 
the generalized-dilation energy $C_6$ when $n=18/5$, 
spanning a 1-dimensional vector space. 
\end{thm}

\subsection{Conservation laws of $u_{tt}=(c+(a+b)u_{r}^{p})u_{rr}+(n-1)(cu_{r}+bu_{r}^{p+1})/r$.}\label{sec:Bconslaws}

\indent\newline\indent
In the following tables we list the kinematic conservation laws \eqref{kinematicT}--\eqref{kinematicQ}
and generalized-energy conservation laws \eqref{genergyQ}--\eqref{energyT}
of the radial wave equation \eqref{WEreducrad2} 
in all cases such that this equation is
nonlinear (i.e.\ $p(p+1)b\neq0$ or $p(a+b)\neq0$), 
hyperbolic (i.e.\ $p(b+a)\neq0$ or $c\neq0$), 
and multi-dimensional (i.e.\ $n\neq1$, $c\neq0$ or $(p+1)b\neq0$). 
These three restrictions are equivalent to the inequality 
$(n-1)p(b^2+c^2)((b+a)^2+b^2c^2(p+1)^2)\neq0$ holding on the nonlinearity power $p$, 
the dimension $n$, and the coefficients $a$, $b$, $c$.
Wherever it is convenient, we will use the notation
\begin{equation}
 k=a+b
\end{equation}
and 
\begin{equation}\label{radpower}
 m=b(n-1)(p+1)/(a+b).
\end{equation}

To begin, 
the kinematic conserved densities, fluxes, and corresponding multipliers 
are listed in Table~\ref{radBconslaw0th}.

\begin{table}
\begin{changemargin}{-1cm}{-1cm}
\begin{center}\refstepcounter{tabul}\label{radBconslaw0th}
      \setcounter{tbn}{0}
      \begin{tabular}{|c|c|c|c|c|}
      \hline\vspacebefore
      \hfill &\hfill $Q\hfill$ &\hfill $T\hfill$ &\hfill $X\hfill$ &\hfill {conditions\hfill}  \\
      \hline\vspacebeforemore
      $1$ 
        & $r^{m}$ 
        & $r^{m}u_{t}$ 
        & $-r^{m}\Big(cu_{r}+\dfrac{b(n-1)}{m}u_{r}^{p+1}\Big)$
        & $c(k-(p+1)b)=0$\\[1.2ex]
      \hline\vspacebeforemore
      $2$ 
        & $r^{m}t$ 
        & $r^{m}(tu_{t}-u)$ 
        & $-r^{m}t\Big(cu_{r}+\dfrac{b(n-1)}{m}u_{r}^{p+1}\Big)$
        & $c(k-(p+1)b)=0$\\[1.2ex]
      \hline\vspacebeforemore
      $3$ 
        & $r$ 
        & $ru_{t}$ 
        & $-b(n-1)ru_{r}^{p+1}-c(ru_{r}+(n-2)u)$
        & $\begin{aligned}&k=(p+1)(n-1)b, c\neq0\\&n\neq2\end{aligned}$ \\[1.2ex]
      \hline\vspacebeforemore
      $4$ 
        & $rt$ 
        & $r(tu_{t}-u)$ 
        & $-t(b(n-1)ru_{r}^{p+1}+c(ru_{r}+(n-2)u))$
        & $\begin{aligned}&k=(p+1)(n-1)b, c\neq0\\&n\neq2\end{aligned}$ \\[1.2ex]
      \hline
      \end{tabular}
      \end{center}
\caption{Kinematic conservation laws for $(n-1)(b^2(p+1)^{2}+c^2)\neq0$, $p\big(k^2+b^2c^2(p+1)^2\big)\neq0$.}
\end{changemargin}
\end{table}

These conservation laws are characterized by three distinguished relations among $b,c,k$:
$k=(p+1)b$ (i.e. $a=pb$), or $c=0$, or $k=(p+1)(n-1)b$ (i.e. $a=(p(n-1)+n-2)b$), $c\neq0$. 
When $k=(p+1)b$ the radial power \eqref{radpower} reduces to $m=n-1$, so thus in this case 
the conserved densities $T_1$ and $T_2$ arise from the $n$-dimensional divergence structure \eqref{divergence} 
for the radial wave equation \eqref{WEreducrad2} with $c\neq0$. 
When $c=0$ the conserved densities $T_1$ and $T_2$ arise from the alternative divergence structure \eqref{altdivergence} for this wave equation \eqref{WEreducrad2},
with an arbitrary radial power \eqref{radpower}. 
In contrast, when $c\neq0$ and $k=(p+1)(n-1)b$, 
the conserved densities $T_3$ and $T_4$ arise from a different divergence structure 
$(ru_t)_t = (rF)_r$ 
where the flux is given by $F=c(r u_r +(n-2)u) +(k/(p+1))u_r^{p+1}$. 
We note that, in the case $c=0$, 
this divergence structure coincides with the alternative divergence structure \eqref{altdivergence}, since the relation $k=(p+1)(n-1)b$ is equivalent to $m=1$. 

We will now summarize the structure of the vector space spanned by the conserved quantities \eqref{C} arising from these kinematic conserved densities. 

\begin{thm}\label{Bkinconslawsvs}
A multi-dimensional, hyperbolic, nonlinear radial wave equation \eqref{WEreducrad2}
admits kinematic conservation laws only in the following cases:
\newline
{\rm (i)}
For $c\neq0$, $k=(p+1)b\neq0$, 
a basis of kinematic conserved quantities is comprised by
\begin{equation}\label{Bkinbasis1}
C_1=\int_0^\infty u_{t}r^{n-1} dr, \quad
C_2=\int_0^\infty (tu_{t}-u)r^{n-1} dr.
\end{equation}
\newline
{\rm (ii)}
For $c\neq0$, $k=(p+1)(n-1)b\neq0$, $n\neq2$, 
a basis of kinematic conserved quantities is comprised by
\begin{equation}\label{Bkinbasis2}
C_3=\int_0^\infty u_{t}r dr, \quad
C_4=\int_0^\infty (tu_{t}-u)r dr.
\end{equation}
{\rm (iii)}
For $c=0$, $k\neq (p+1)b$, 
a basis of kinematic conserved quantities is comprised by
\begin{equation}\label{Bkinbasis3}
C_1=\int_0^\infty u_{t}r^{m} dr, \quad
C_2=\int_0^\infty (tu_{t}-u)r^{m} dr.
\end{equation}
\newline
In all cases the conserved quantities span a 2-dimensional vector space. 
\end{thm}

The conditions on the coefficients $b,c,k=a+b$ coming from Theorem~\ref{Bkinconslawsvs}
combined with Proposition~\ref{structures}
imply the following interesting result. 

\begin{cor}
A multi-dimensional, hyperbolic, nonlinear radial wave equation \eqref{WEreducrad2}
possesses kinematic conservation laws iff it has a quasilinear divergence structure. 
\end{cor}

Next, the generalized-energy conserved densities and fluxes are listed in Table~\ref{radBconslaw1st};
corresponding multipliers are listed in Table~\ref{radBmultiplier1st}.

\begin{table}
\begin{changemargin}{-2cm}{-1cm}
 {\begin{center}\refstepcounter{tabul}\label{radBconslaw1st}
      \setcounter{tbn}{0}
      \begin{tabular}{|c|c|c|c|c|}
      \hline\vspacebefore
      \hfill &\hfill $T\hfill$ &\hfill $X\hfill$ &\hfill {conditions\hfill} \\
      \hline\vspacebeforemore
      $1$ 
        & $r^{n-1}(\tfrac{1}{2}(u_{t}^{2}+cu_{r}^{2})+bh(u_{r}))$ 
	& $-r^{n-1}(cu_{r}+bu_{r}^{p+1})u_{t} $
        & $k=(p+1)b$, $c\neq0$ \\[1.2ex]
      \hline\vspacebeforemore
      $2$ 
        & $\begin{aligned}r^{-2}\big(&\tfrac{1}{2}t(u_{t}^2+cu_{r}^{2}+2bh(u_{r}))\\
	    &+(ru_{r}-u)u_{t}\big)
	   \end{aligned}$  
	& $\begin{aligned}-\tfrac{1}{2}r^{-2}\big(&r(u_{t}^{2}+cu_{r}^2+2kh(u_{r}))\\
	    &+2(c+bu_{r}^{p})(tu_{t}-u)u_{r}\big) 
	   \end{aligned}$
        & $\begin{aligned}& k=(p+1)b, c\neq0 \\& n=-1 \end{aligned}$ \\[1.2ex]
      \hline\vspacebeforemore
      $3$ 
        & $r^{m}\big(\tfrac{1}{2}u_{t}^2+\dfrac{b(n-1)}{m}h(u_{r})\big)$ 
	& $-r^{m}\big(ku_{t}h(u_{r})\big) $
        & $c=0$ \\[1.2ex]
      \hline\vspacebeforemore
      $4$ 
        & $\begin{aligned}\tfrac{1}{2}r^{m}\big(&(p(m+3)+4)t\big(u_{t}^{2}+\dfrac{2b(n-1)}{m}h(u_{r})\big)\\
            &+2\big((p+4)ru_{r}+(2m-p)u\big)u_{t}\big)
           \end{aligned}$ 
	& $\begin{aligned}-\tfrac{1}{2}r^m\big(&(p+4)r(u_{t}^{2}+2kh(u_{r}))\\
	    &+2\big((p(m+3)+4)tu_{t}\\&+(2m-p)u\big)\dfrac{b(n-1)}{m}u_{r}^{p+1}\big) 
	   \end{aligned}$
        & $\begin{aligned}&c=0\\& p\neq-4 \end{aligned}$ \\[1.2ex]
      \hline\vspacebeforemore
      $5$ 
        & $\tfrac{1}{3}r^{m}\big(3(tu_{t}-u)u_{t}+ktu_{r}^{-2}\big) $ 
	& $\tfrac{1}{3}r^{m}\big(k(2tu_{t}-u)u_{r}^{-3}\big) $
        & $\begin{aligned}&c=0, \\& p=-4 \end{aligned}$ \\[1.2ex]
      \hline\vspacebeforemore
      $6$ 
        & $\tfrac{1}{6}r^{m}\big(3(tu_{t}-u)^{2}+kt^2 u_r^{-2}\big) $
	& $\tfrac{1}{3}r^{m}\big(kt(tu_{t}-u)u_{r}^{-3}\big) $
        & $\begin{aligned}&c=0 \\& p=-4 \end{aligned}$ \\[1.2ex]
      \hline\vspacebeforemore
      $7$  
        & $u_{r}u_{t}$ 
	& $-\tfrac{1}{2}(u_{t}^{2}+2k\ln{u_{r}})-(n-1)b\ln{r} $
        & $\begin{aligned}&c=0,\\ &p=-2\end{aligned} $ \\[1.2ex]
      \hline\vspacebeforemore
      $8$ 
        & $\tfrac{1}{2}r^{-2}\big(u_{t}(2ru_{r}-u)\big) $ 
	& $\begin{aligned}-\tfrac{1}{2}r^{-2}\big(&b(n-1)uu_{r}^{-3}\\
	     &+r(2u_{t}^{2}-3b(n-1)u_{r}^{-2})\big)    
	   \end{aligned}$
        & $\begin{aligned}&c=0,\\&k=3b(n-1)/2,\\&p=-4 \end{aligned} $ \\[1.2ex]
      \hline\vspacebeforemore
      $9$ 
        & $\begin{aligned}\tfrac{1}{2}r^{-1}\big(&(2u-tu_{t})u_{t}+2(r\ln{r})u_{r}u_{t}\\
	    &+2kt\ln{u_{r}}\big) 
	   \end{aligned}$ 
	& $\begin{aligned}-\tfrac{1}{2}\big(&(k(2(1+\ln{u_{r}})+\ln{r})+u_{t}^{2})\ln{r}\\
	      &+2k(tu_{t}-u)u_{r}^{-1}r^{-1}\big)\end{aligned} $
        & $\begin{aligned}&c=0,\\& k=b(n-1),\\& p=-2 \end{aligned}$ \\[1.2ex]
      \hline\vspacebeforemore
      $10$ 
        & $(\tfrac{1}{2}u_{t}^2-k(1-\ln{u_{r}}))u_{r}-b(n-1)u/r $ 
	& $-\tfrac{1}{6}(u_{t}^{2}+6k\ln{u_{r}})u_{t} $
        & $\begin{aligned}&c=0,\\& p=-2\end{aligned}$ \\[1.2ex]
      \hline\vspacebeforemore
      $11$ 
        & $\begin{aligned}\tfrac{1}{2}r^{m}\Big(&u\big(u_{t}^{2}+\dfrac{2b(n-1)}{m}h(u_{r})\big)\\
           &-\dfrac{p-2}{p+2}r(2k\displaystyle\int h(u_{r})du_{r}+u_{r}u_{t}^{2})\Big)\end{aligned}$ 
	& $\begin{aligned}r^{m}\big(&\dfrac{p-2}{p+2}r(\tfrac{1}{6}u_{t}^{2}+kh(u_{r}))u_{t}\\
	    &-\dfrac{b(n-1)}{m}uu_{t}u_{r}^{p+1}\big) \end{aligned}$
        & $\begin{aligned}&c=0,\\ &4k=b(1-n)(p-2),\\ &p\neq-2\end{aligned}$ \\[1.2ex]
      \hline\vspacebeforemore
      $12$ 
        & $\begin{aligned}\tfrac{1}{4}r^{-2}\big(&2(2u-tu_{t})(u-3ru_{r})u_{t}\\
              &+bt(n-1)(9ru_{r}-u)u_{r}^{-2}\big) \end{aligned}$
	& $\begin{aligned}\tfrac{1}{4}r^{-2}\big(&b(n-1)(9ru_{r}-2u)(tu_{t}-u)u_{r}^{-3}\\
	    &-2ru_{t}^{2}(tu_{t}-3u)\big)\end{aligned} $
        & $\begin{aligned}&c=0, \\& 4k=b(1-n)(p-2),\\& p=-4 \end{aligned}$ \\[1.2ex]
      \hline
      \end{tabular}
      \end{center}}
\caption{Generalized-energy conservation laws for $(n-1)(b^2(p+1)^{2}+c^2)\neq0$, $p\big(k^2+b^2c^2(p+1)^2\big)\neq0$.}
\end{changemargin}
\end{table}

\begin{table}
\begin{changemargin}{-2cm}{-1cm}
 {\begin{center}\refstepcounter{tabul}\label{radBmultiplier1st}
      \setcounter{tbn}{0}
      \begin{tabular}{|c|c|c|c|c|}
      \hline\vspacebefore
      \hfill &\hfill $Q\hfill$ &\hfill {conditions\hfill} \\
      \hline\vspacebeforemore
      $1$ 
        &$r^{n-1}u_{t}$ 
        & $k=(p+1)b$, $c\neq0$ \\[1.2ex]
      \hline\vspacebeforemore
      $2$ 
        & $r^{-2}(tu_{t}+ru_{r}-u)$ 
        & $\begin{aligned}& k=(p+1)b, c\neq0 \\& n=-1 \end{aligned}$ \\[1.2ex]
      \hline\vspacebeforemore
      $3$ 
        & $r^{m}u_{t}$ 
        & $c=0$ \\[1.2ex]
      \hline\vspacebeforemore
      $4$ 
        & $\begin{aligned}r^{m}\Big(&\big(p(m+3)+4\big)tu_{t}\\ 
            &+(p+4)ru_{r}+(2m-p)u\Big)
          \end{aligned}$ 
        & $\begin{aligned}&c=0\\&p\neq-4 \end{aligned}$ \\[1.2ex]
      \hline\vspacebeforemore
      $5$ 
        & $r^{m}(2tu_{t}-u)$ 
        & $\begin{aligned}&c=0, \\& p=-4 \end{aligned}$ \\[1.2ex]
      \hline\vspacebeforemore
      $6$ 
        & $r^{m}t(tu_{t}-u)$ 
        & $\begin{aligned}&c=0 \\& p=-4 \end{aligned}$ \\[1.2ex]
      \hline\vspacebeforemore
      $7$  
        & $u_{r} $ 
        & $\begin{aligned}&c=0,\\ &p=-2 \end{aligned}$ \\[1.2ex]
      \hline\vspacebeforemore
      $8$ 
        & $r^{-2}(2ru_{r}-u)$  
        & $\begin{aligned}&c=0,\\&k=3b(n-1)/2,\\&p=-4 \end{aligned} $ \\[1.2ex]
      \hline\vspacebeforemore
      $9$ 
        & $r^{-1}(r(\ln{r})u_{r}-tu_{t}+u) $ 
        & $\begin{aligned}&c=0,\\& k=b(n-1),\\& p=-2 \end{aligned}$ \\[1.2ex]
      \hline\vspacebeforemore
      $10$ 
        & $u_{r}u_{t}$ 
        & $\begin{aligned}&c=0,\\& p=-2\end{aligned}$ \\[1.2ex]
      \hline\vspacebeforemore
      $11$ 
        & $\begin{aligned}&r^{-4(p+1)/(p-2)}\big(u-\dfrac{p-2}{p+2}ru_{r}\big)u_{t} \end{aligned}$ 
        & $\begin{aligned}&c=0,\\ &4k=b(1-n)(p-2),\\ &p\neq-2\end{aligned}$ \\[1.2ex]
      \hline\vspacebeforemore
      $12$ 
        & $r^{-2}(3ru_{r}-u)(tu_{t}-u) $ 
        & $\begin{aligned}&c=0, \\& 4k=b(1-n)(p-2),\\& p=-4 \end{aligned}$ \\[1.2ex]
      \hline
      \end{tabular}
      \end{center}}
\caption{Generalized-energy conservation laws for $(n-1)(b^{2}(p+1)^{2}+c^2)\neq0$, $p\big((b+a)^2+b^2c^2(p+1)^2\big)\neq0$.}
\end{changemargin}
\end{table}

We see from Table~\ref{radBconslaw1st} that these conservation laws are characterized by two distinguished
relations among the coefficients $b$, $c$, $k$: $c=0$, or $k=(p+1)b$, $c\neq0$.
These relations are simply the conditions for a radial wave equation \eqref{WEreducrad2} to possess an 
$n$-dimensional variational structure \eqref{variational} or an alternative variational structure \eqref{altvariational}
based on the radial power \eqref{scalcond1}.
Thus, all generalized-energy conservation laws for this wave equation correspond to variational symmetries through Noether's
theorem.
We will summarize this correspondence in the next subsection~\ref{sec:BNoether}.
In contrast to the situation for the radial wave equation \eqref{WEreducrad1}, some of the symmetries here are of contact type.
Interestingly, all of the generalized-energy multipliers corresponding to the contact symmetries have a product form
distinguished by the relations $b=0$, or $p=-2$, or $m=4(p+1)/(2-p)$.
We will show later in subsection \ref{sec:BMorawetz} that these multipliers for the radial wave equation \eqref{WEreducrad2}
can be derived alternatively from a variational type of Morawetz radial identity
which produces conservation laws from symmetries  
in a similar way to the generalized-energy conservation laws that arise in the non-variational case of the radial wave
equation \eqref{WEreducrad1} discussed in subsection \eqref{sec:AMorawetz}.

\subsubsection{Noether correspondence for $u_{tt}=\big(c+b(p+1)u_{r}^{p}\big)u_{rr} +(n-1)(cu_{r}+bu_{r}^{p+1})/r$ and
            $u_{tt}=ku_{r}^{p}u_{rr}+(n-1)bu_{r}^{p+1}/r$.}\label{sec:BNoether}

In the two variational cases, $k=(p+1)b$ and $c=0$,  
multipliers for conservation laws of the radial wave equation \eqref{WEreducrad2}
are in one-to-one correspondence to variational symmetries of the respective Lagrangians \eqref{Lagrang2} and 
\eqref{altvariational}. 
These cases can be merged if we consider the Lagrangian functional
\begin{equation}\label{mergLagrang}
{\mathcal L}=\int_{t_0}^{t_1}\int_{0}^{\infty}Lr^{m}\;drdt ,\quad
L=\tfrac{1}{2}(-u_{t}^{2}+cu_{r}^{2})+\int \wt{F}(r,u_{r})du_{r}
\end{equation}
given in terms of the expression \eqref{altdivergence} for $\wt{F}$, 
with the coefficients $b$, $c$, $k$ satisfying the condition
\begin{equation}\label{mergcond}
 c(k-(p+1)b)=0.
\end{equation}
Then Noether's theorem states that 
\begin{equation}
 {\rm X}=P\partial/\partial u
\end{equation}
is a variational symmetry satisfying
\begin{equation}
 r^{m}\pr{\rm X}L=D_{t}A+D_{r}B
\end{equation}
for some scalar functions $A$ and $B$ depending on $t$, $r$, $u$, $u_{t}$, and $r$-derivatives of $u$, $u_{t}$,
iff 
\begin{equation}
\frac{\delta {\mathcal L}}{\delta u}P
= (u_{tt}-gu_{rr}-f)r^{m}P = D_{t}T +D_{r}X
\end{equation}
is the characteristic form for a conservation law given by the conserved density and flux 
\begin{equation}
T=A-Q\frac{\partial L}{\partial u_{t}},\quad 
X=B-Q\frac{\partial L}{\partial u_{r}}
\end{equation}
where
\begin{equation}
 Q=r^{m}P
\end{equation}
is the multiplier.

For the generalized-energy multipliers in Table~\ref{radBmultiplier1st}, we see that either $Q$ is linear in $u_t$ and $u_r$,
\begin{equation}\label{notherQpoint}
 Q=u_{t}\alpha_{0}(t,r)+u_{r}\beta_{1}(t,r)+u\beta_{0}(t,r)
\end{equation}
where $\alpha_{0}$, $\beta_{1}$, $\beta_{0}$ have no dependence on $u$,
in which case the corresponding variational symmetry is a restricted type of point symmetry, 
or $Q$ involves a product of $u_{t}$ and $u_{r}$, 
\begin{equation}\label{notherQcontact}
 Q=u_{t}\big(u_{r}\alpha_{1}(t,r)+\alpha_{0}(t,r,u)\big)+u_{r}\beta_{1}(t,r,u)+\beta_{0}(t,r,u)
\end{equation}
where $\alpha_{1}$ has no dependence on $u$, while $\alpha_{0}$ and $\beta_{1}$ are at most linear in $u$, 
and $\beta_{0}$ is at most quadratic in $u$, 
in which case the corresponding variational symmetry ${\rm X}$ is a restricted type of contact symmetry. 
This correspondence is summarized in Tables~\ref{radBgeneratorspoint} and \ref{radBgeneratorscontact}, 
where $C$ denotes the conserved quantity \eqref{C} given by the conservation laws, 
and $\hat{\rm X}$ denotes the variational symmetry generator given by the canonical form 
\begin{equation}\label{notherXpoint}
 \hat{{\rm X}}=r^{-m}(-\alpha_{0}\partial/\partial t-\beta_{1}\partial/\partial r+\beta_{0}\partial/\partial u)
\end{equation}
in the point case \eqref{notherQpoint}, 
or by
\begin{equation}\label{notherXcontact}
\begin{aligned}
 \hat{{\rm X}}=r^{-m}\bigg(&-(u_{r}\alpha_{1}+\alpha_{0})\partial/\partial t-(u_{t}\alpha_{1}+\beta_{1})\partial/\partial r
   +\beta_{0}\partial/\partial u\\
   &+\Big(u_{t}^{2}\dfrac{\partial\alpha_{0}}{\partial u}
     +u_{t}u_{r}\big(\dfrac{\partial\alpha_{1}}{\partial t}+\dfrac{\partial\beta_{1}}{\partial u}\big)
     +u_{t}\big(\dfrac{\partial\alpha_{0}}{\partial t}+\dfrac{\partial\beta_{0}}{\partial u}\big)
     +u_{r}\dfrac{\partial\beta_{1}}{\partial t}+\dfrac{\partial\beta_{0}}{\partial t}\Big)\partial/\partial u_{t}\\
   &+\Big(u_{r}^{2}\dfrac{\partial\beta_{1}}{\partial u}
     +u_{t}u_{r}\big(\dfrac{\partial\alpha_{1}}{\partial r}+\dfrac{\partial\alpha_{0}}{\partial u}\big)
     +u_{t}\dfrac{\partial\alpha_{0}}{\partial r}
     +u_{r}\big(\dfrac{\partial\beta_{1}}{\partial r}+\dfrac{\partial\beta_{0}}{\partial u}\big)
     +\dfrac{\partial\beta_{0}}{\partial r}\Big)\partial/\partial u_{t}\bigg)
\end{aligned}
\end{equation}
in the contact case \eqref{notherQcontact}.

\begin{table} 
\begin{changemargin}{-2cm}{-1cm}
 {\begin{center}\refstepcounter{tabul}\label{radBgeneratorspoint}
      \setcounter{tbn}{0}
      \begin{tabular}{|c|c|c|c|c|}
      \hline\vspacebefore
      \hfill &\hfill $\hat{\rm X}\hfill$ &\hfill $C\hfill$ &\hfill {conditions\hfill} \\
      \hline\vspacebeforemore
      $1\ \&\ 3$ 
	& $-\partial/\partial t$
        & $\displaystyle\int_{0}^{\infty}(\tfrac{1}{2}(u_{t}^{2}+cu_{r}^{2})+\dfrac{b(n-1)}{m}h(u_{r}))r^{n-1} \;dr$ 
        & $c(k-(p+1)b)=0$ \\[1.2ex]
      \hline\vspacebeforemore
      $2$ 
	& $\begin{aligned}&-t\partial/\partial t-r\partial/\partial r\\&-u\partial/\partial u\end{aligned}$
        & $\displaystyle\int_{0}^{\infty}\big(\tfrac{1}{2}t(u_{t}^2+cu_{r}^{2}+2bh(u_{r}))+(ru_{r}-u)u_{t}\big)r^{-2} \;dr$  
        & $\begin{aligned}& k=(p+1)b, c\neq0 \\& n=-1 \end{aligned}$ \\[1.2ex]
      \hline\vspacebeforemore
      $4$ 
	& $\begin{aligned}&-\big(p(m+3)+4\big)t\partial/\partial t\\
            &-(p+4)r\partial/\partial r\\
            &+(2m-p)u\partial/\partial u
          \end{aligned}$
        & $\begin{aligned}\displaystyle\int_{0}^{\infty}\tfrac{1}{2}\big(&(p(m+3)+4)t\big(u_{t}^{2}+\dfrac{2b(n-1)}{m}h(u_{r})\big)\\
            &+2\big((p+4)ru_{r}+(2m-p)u\big)u_{t}\big)r^{m} \;dr
           \end{aligned}$ 
        & $\begin{aligned}&c=0\\& p\neq-4 \end{aligned}$ \\[1.2ex]
      \hline\vspacebeforemore
      $5$ 
	& $-2t\partial/\partial t-u\partial/\partial u$
        & $\displaystyle\int_{0}^{\infty}\tfrac{1}{3}\big(3(tu_{t}-u)u_{t}+ktu_{r}^{-2}\big)r^{m} \;dr$ 
        & $\begin{aligned}&c=0, \\& p=-4 \end{aligned}$ \\[1.2ex]
      \hline\vspacebeforemore
      $6$ 
	& $-t^{2}\partial/\partial t-tu\partial/\partial u$
        & $\displaystyle\int_{0}^{\infty}\tfrac{1}{6}\big(3(tu_{t}-u)^{2}+kt^2 u_r^{-2}\big)r^{m} \;dr$
        & $\begin{aligned}&c=0 \\& p=-4 \end{aligned}$ \\[1.2ex]
      \hline\vspacebeforemore
      $7$  
	& $-r^{-m}\partial/\partial r$
        & $\displaystyle\int_{0}^{\infty}u_{r}u_{t}\;dr$ 
        & $\begin{aligned}&c=0,\\ &p=-2\end{aligned} $ \\[1.2ex]
      \hline\vspacebeforemore
      $8$ 
	& $-2r\partial/\partial r-u\partial/\partial u$
        & $\displaystyle\int_{0}^{\infty}u_{t}\big(2ru_{r}-u\big)r^{-2} \;dr$ 
        & $\begin{aligned}&c=0,\\&k=3b(n-1)/2,\\&p=-4 \end{aligned} $ \\[1.2ex]
      \hline\vspacebeforemore
      $9$ 
	& $\begin{aligned}&t\partial/\partial t-r\ln{r}\partial/\partial r\\&+u\partial/\partial u\end{aligned}$
        & $\begin{aligned}\displaystyle\int_{0}^{\infty}\tfrac{1}{2}\big(&(2u-tu_{t})u_{t}+2(r\ln{r})u_{r}u_{t}\\
	    &+2kt\ln{u_{r}}\big)r^{-1} \;dr\end{aligned}$ 
        & $\begin{aligned}&c=0,\ k=b(n-1),\\& p=-2 \end{aligned}$ \\[1.2ex]
      \hline
      \end{tabular}
      \end{center}}
\caption{Generalized-energy conserved quantities and variational point symmetries for $(n-1)(b^2(p+1)^{2}+c^2)\neq0$, $p\big(k^2+b^2c^2(p+1)^2\big)\neq0$.}
\end{changemargin}
\end{table}

Conserved quantity $C_1(=C_{3})$ is an energy arising from the time-translation symmetry $\hatrmX{1}(=\hatrmX{3})$. 
Conserved quantities $C_2$, $C_{4}$ and $C_{5}$ are similarity energies arising from the scaling symmetries $\hatrmX{2}$,
$\hatrmX{4}$ and $\hatrmX{5}$, 
while conserved quantity $C_9$ is a logarithmic counterpart of $C_4$. 
Conserved quantity $C_8$ and $C_{7}$ are radial momenta arising respectively from the radial scaling symmetry $\hatrmX{8}$ and the
non-rigid dilation symmetry $\hatrmX{7}$. 
Conserved quantity $C_6$ is a conformal energy arising from the temporal inversion symmetry $\hatrmX{6}$. 

We will now state the vector space structure spanned by 
these conserved quantities as well as the algebra structure 
determined by the corresponding variational symmetries. 

\begin{thm}\label{Bnoetherconslawsvs}
For a multi-dimensional, hyperbolic, nonlinear radial wave equation \eqref{WEreducrad2}
that has a variational structure \eqref{variational} when $a=pb\neq0$, $c\neq0$, and an alternative
variational structure \eqref{altvariational} when $c=0$, 
variational point symmetries are admitted in the following cases, 
corresponding to all generalized-energy conservation laws whose multipliers are at most linear in derivatives. 
\newline
{\rm (i)}
For $c=0$, the variational point symmetries comprise 
the time-translation $\hatrmX{3}$;
the scaling symmetry $\hatrmX{4}$; 
the temporal scaling and inversion symmetries $\hatrmX{5}$ and $\hatrmX{6}$ when $p=-4$;
the radial scaling symmetry $\hatrmX{8}$ when $p=-4$, $k=3b(n-1)/2$;
the non-rigid dilation symmetries $\hatrmX{7}$ when $p=-2$
and $\hatrmX{9}$ when $p=-2$, $k=b(n-1)$.   
In the subcase $p\neq-2$, $p\neq-4$,
the time-translation $\hatrmX{3}$ and the scaling $\hatrmX{4}$ generate 
a two-dimensional algebra with the commutator structure 
\begin{equation}\label{Bnoethercom3}
[\hatrmX{3},\hatrmX{4}]=-(p(m+3)+4)\hatrmX{3}.
\end{equation}
The corresponding two-dimensional vector space of conserved quantities is spanned by 
the energy $C_3$ and the similarity energy $C_4$. 
In the subcase $p=-4$, $k\neq 3b(n-1)/2$, 
the scaling $\hatrmX{5}$ and temporal inversion $\hatrmX{6}$
enlarge the commutator structure \eqref{Bnoethercom3} by 
\begin{equation}\label{Bnoethercom7}
[\hatrmX{3},\hatrmX{5}]=-2\hatrmX{3},
[\hatrmX{4},\hatrmX{5}]=0,
[\hatrmX{3},\hatrmX{6}]=\hatrmX{5},
[\hatrmX{4},\hatrmX{6}]=4(m+2)\hatrmX{6}
\end{equation}
generating a four-dimensional algebra whose corresponding vector space of conserved quantities is spanned by 
$C_3$, $C_4$, and the similarity energy $C_5$ in addition to the conformal energy $C_{6}$. 
In the subcase $p=-4$, $k=3b(n-1)/2$, 
the radial scaling $\hatrmX{8}$ 
enlarges the commutator structure \eqref{Bnoethercom3} and \eqref{Bnoethercom7} by 
\begin{equation}\label{Bnoethercom8}
[\hatrmX{3},\hatrmX{8}]=[\hatrmX{4},\hatrmX{8}]=[\hatrmX{5},\hatrmX{8}]=[\hatrmX{6},\hatrmX{8}]=0
\end{equation}
generating a five-dimensional algebra whose corresponding vector space of conserved quantities is spanned by 
$C_3$, $C_4$, $C_5$, $C_6$, and $C_8$. 
In the subcase $p=-2$, $k\neq b(n-1)$, 
the non-rigid dilation $\hatrmX{7}$ enlarges the commutator structure \eqref{Bnoethercom3} by 
\begin{equation}\label{Bnoethercom5}
[\hatrmX{3},\hatrmX{7}]=0,
[\hatrmX{4},\hatrmX{7}]=2(m+1)\hatrmX{7}
\end{equation}
generating another three-dimensional algebra, whose corresponding vector space of conserved quantities is spanned by 
$C_3$, $C_4$ and the radial momentum $C_7$. 
In the subcase $p=-2$, $k=b(n-1)$, 
the non-rigid dilation symmetry $\hatrmX{9}$ 
enlarges the commutator structure \eqref{Bnoethercom3} and \eqref{Bnoethercom5} by 
\begin{equation}\label{Bnoethercom6}
[\hatrmX{3},\hatrmX{9}]=\hatrmX{3},
[\hatrmX{4},\hatrmX{9}]=-2\hatrmX{7},
[\hatrmX{7},\hatrmX{9}]=-\hatrmX{7}
\end{equation}
generating a five-dimensional algebra whose corresponding vector space of conserved quantities is spanned by 
$C_3$, $C_4$, $C_7$, and the logarithmic similarity energy $C_{9}$. 
\newline
{\rm (ii)}
For $c\neq0$, $k=(p+1)b\neq0$, the variational point symmetries comprise 
the time-translation $\hatrmX{1}$, and the scaling $\hatrmX{2}$ when $n=-1$.
In the subcase $n\neq-1$, the time-translation $\hatrmX{1}$ generates 
a one-dimensional algebra,
while the corresponding vector space of conserved quantities is spanned by 
the energy $C_1$. 
In the subcase $n=-1$, 
the scaling $\hatrmX{2}$ and the time-translation $\hatrmX{1}$ generate a two-dimensional algebra with commutator structure  
\begin{equation}\label{Bnoethercom9}
[\hatrmX{1},\hatrmX{2}]=-\hatrmX{1}
\end{equation}
while the corresponding vector space of conserved quantities is spanned by 
$C_1$ and the similarity energy $C_2$. 
\end{thm}

\begin{table}
\begin{changemargin}{-2cm}{-1cm}
 {\begin{center}\refstepcounter{tabul}\label{radBgeneratorscontact}
      \setcounter{tbn}{0}
      \begin{tabular}{|c|c|c|c|c|}
      \hline\vspacebefore
      \hfill &\hfill $\hat{\rm X}\hfill$ &\hfill $C\hfill$ &\hfill {conditions\hfill} \\
      \hline\vspacebeforemore
      $10$ 
	& $\begin{aligned}-r^{-m}(& u_{r}\partial/\partial t+u_{t}\partial/\partial r\\
	    &+u_{r}u_{t}\partial/\partial u\\
	    &+mr^{-1}u_{r}u_{t}\partial/\partial u_{r})\end{aligned}$
        & $\displaystyle\int_{0}^{\infty}(\tfrac{1}{2}u_{t}^2-k(1-\ln{u_{r}}))u_{r}-b(n-1)u/r \;dr$ 
        & $\begin{aligned}&c=0,\\& p=-2\end{aligned}$ \\[1.2ex]
      \hline\vspacebeforemore
      $11$ 
	& $\begin{aligned}&((p-2)u_{r}-(p+2)u)\partial/\partial t\\
	    &+(p-2)u_{t}\partial/\partial r\\
	    &+(p-2)u_{r}u_{t}\partial/\partial u\\
	    &+(p+2)u_{t}^{2}\partial/\partial u_{t}\\
	    &+(p+2)u_{r}u_{t}\partial/\partial u_{r}\end{aligned}$
        & $\begin{aligned}\displaystyle\int_{0}^{\infty}\Big(&u\big(u_{t}^{2}+\dfrac{2b(n-1)}{m}h(u_{r})\big)\\
           &-\dfrac{p-2}{p+2}r(2k\displaystyle\int h(u_{r})du_{r}+u_{r}u_{t}^{2})\Big)r^{m} \;dr\end{aligned}$ 
        & $\begin{aligned}&c=0,\\ &4k=-b(p-2)(n-1),\\ &p\neq-2\end{aligned}$ \\[1.2ex]
      \hline\vspacebeforemore
      $12$ 
	& $\begin{aligned}&(u-3ru_{r})\partial/\partial t\\
	    &+3r(u-tu_{t})\partial/\partial r\\
	    &+(u^{2}-3rtu_{r}u_{t})\partial/\partial u\\
	    &+(u-tu_{t})u_{t}\partial/\partial u_{t}\\
	    &+(2tu_{t}-3ru_{r}-u)u_{r}\partial/\partial u_{r}\end{aligned}$
        & $\begin{aligned}\displaystyle\int_{0}^{\infty}\tfrac{1}{4}\big(&2(2u-tu_{t})(u-3ru_{r})u_{t}\\
              &+bt(n-1)(9ru_{r}-u)u_{r}^{-2}\big)r^{-2}\;dr\end{aligned}$
        & $\begin{aligned}&c=0, \\& 4k=-b(p-2)(n-1),\\& p=-4 \end{aligned}$ \\[1.2ex]
      \hline
      \end{tabular}
      \end{center}}
\caption{Generalized-energy conserved quantities and variational contact symmetries for $(n-1)b(p+1)\neq0$, $pk\neq0$.}
\end{changemargin}
\end{table}

We will now state the vector space structure spanned by 
these generalized-energy conserved quantities as well as the algebra structure 
determined by the corresponding variational contact symmetries. 

\begin{thm}\label{Baltnoetherconslawsvs}
For a multi-dimensional, hyperbolic, nonlinear radial wave equation \eqref{WEreducrad2}
that has a variational structure \eqref{altvariational} when $c=0$, 
variational contact symmetries are admitted in the following cases, 
corresponding to all generalized-energy conservation laws whose multipliers are nonlinear in derivatives. 
\newline
{\rm (i)}
For $p=-2$, the variational contact symmetries comprise $\hatrmX{10}$ generating a one-dimensional
algebra, while the corresponding vector space of conserved quantities is spanned by the generalized-energy $C_{10}$.
\newline
{\rm (ii)}
For $p\neq-2$, $4k=-b(n-1)(p-2)$, the variational contact symmetries comprise $\hatrmX{11}$ and $\hatrmX{12}$. 
In the subcase $p\neq-4$, the symmetry $\hatrmX{11}$  generates 
a one-dimensional algebra,
while the corresponding vector space of conserved quantities is spanned by 
the generalized-energy $C_{11}$. 
In the subcase $p=-4$, 
the symmetries $\hatrmX{12}$ and $\hatrmX{11}$ generate a two-dimensional algebra with the commutator structure  
\begin{equation}\label{Bnoethercomc9}
[\hatrmX{11},\hatrmX{12}]=0
\end{equation}
while the corresponding vector space of conserved quantities is spanned by 
generalized-energies $C_{12}$ and $C_{11}$. 
\end{thm}

\subsubsection{Morawetz-type variational identities for 
$u_{tt}=(c+ku_{r}^{p})u_{rr}+(n-1)(cu_{r}+bu_{r}^{p+1})/r$.}\label{sec:BMorawetz}
We will now derive two different Morawetz identities of variational type for the radial wave equation \eqref{WEreducrad2} 
in the case $c=0$ with $k\neq0$, $b(p+1)\neq0$.
The steps are similar to the derivation of the identity \eqref{morawetz1} for the non-variational case of the radial wave 
equation \eqref{WEreducrad1}, and so we omit most details.
 
To derive the first identity, we use the Lagrangian \eqref{altvariational} to write 
\begin{equation}\label{WELrad2}
r^{m}(u_{tt}-ku_{r}^pu_{rr}-(n-1)bu_{r}^{p+1}/r)
= E_u(r^{m} L_{0}) + E_{u}(r^{m}L_{1})
\end{equation}
where $E_u$ is the 1st order Euler-Lagrange operator \eqref{EuLoperator} (i.e. local variational derivative)
with respect to $u$, and where 
\begin{equation}\label{splitL2} 
L_0= -\tfrac{1}{2}u_{t}^2 ,\quad
L_1= \dfrac{k}{p+1}h(u_{r})
\end{equation} 
are the temporal and radial parts of the Lagrangian $L=L_{0}+L_{1}$.
We now multiply equation \eqref{WELrad2} by the expression
\begin{equation}\label{BWcharfunction}
 W=ru_r-\lambda u
\end{equation}
which represents the characteristic function 
given by the generator of a scaling transformation on $(r,u)$, namely 
$\delta t=0$, $\delta r=r$, $\delta u=\lambda u$, 
for some constant $\lambda$. 
By using the operator identity \eqref{operid} combined with the radial scaling properties of 
$L_{0}$ and $L_{1}$, we obtain
\begin{equation}\label{L0term2}
(ru_r-\lambda u)E_u(r^{m} L_0)
= -E_u(r^{m} {\tilde L}_0) +r^{m}(2rD_r+(m-\lambda+1))L_0
\end{equation}
and 
\begin{equation}\label{L1term2}
(ru_r-\lambda u)E_u(r^{m} L_1)
= -E_u(r^{m} {\tilde L}_1) +r^{m}(p-m+1-\lambda(p+1))L_1  
\end{equation}
with
\begin{equation}\label{L0L1new2}
 {\tilde L}_0 = \tfrac{1}{2}(ru_r-\lambda u)u_{t}^{2}
\end{equation}
and
\begin{equation}
 {\tilde L}_1 = \dfrac{k}{p+1}(2r\int h(u_{r})du_{r}+(\lambda u-ru_{r})h(u_{r}))+\wt{\theta}ru_{r}=
  \dfrac{k}{p+3}(ru_{r}h(u_{r})+\lambda uh(u_{r}))+(\wt{\theta}+2)ru_{r}
\end{equation}
where $\wt{\theta}=(\lambda-1)/(m+1)$ when $p=-2$ and $\wt{\theta}=0$ otherwise.
Finally, we require that the terms in equation \eqref{L1term2} containing $L_1$ vanish. 
This yields
\begin{equation}\label{conds2}
 \lambda=1-(n-1)b/k.
\end{equation}
Thus, we have expressed the Euler-Lagrange terms \eqref{L0term2} and \eqref{L1term2} as the sum of a total variational 
derivative $E_u(r^{m} \tilde L)$
for a new Lagrangian 
\begin{equation}\label{Lnew2}
{\tilde L} = {\tilde L}_0 + {\tilde L}_1 ,
\end{equation}
plus a dilational derivative $(rD_r+\tilde\lambda)(r^{m} L_{0}/2)$ with
\begin{equation}\label{constnew2}
\tilde\lambda= (1-m-\lambda)/2=b(n-1)(1-p/2)/k.
\end{equation}
As a result, equation \eqref{WELrad2} multiplied by expression \eqref{BWcharfunction} yields the variational dilation identity
\begin{equation}\label{morawetz2}
r^{m}(ru_r-\lambda u)\big(u_{tt}-ku_{r}^pu_{rr}-(n-1)bu_{r}^{p+2}/r\big)
= -E_u(r^{m}{\tilde L}) -(rD_r+\tilde\lambda)(r^{m}u_t^2).
\end{equation}
This identity \eqref{morawetz2} is able to produce conservation laws 
from multipliers that are given by certain symmetries 
${\rm X} = P\partial/\partial u$ in characteristic form, 
in a manner similar to Noether's theorem \eqref{noetherid}--\eqref{noetherthm}. 
In particular we consider the time translation 
\begin{equation}\label{Xtimetrans}
{\rm X} = u_t\partial/\partial u.
\end{equation}
When the identity \eqref{morawetz2} is multiplied by 
\begin{equation}\label{symmchar2} 
 P = u_t
\end{equation} 
we see first that the Euler-Lagrange term yields a total time derivative
\begin{equation}
-u_t E_u(r^{m}{\tilde L}) 
= D_t\big( r^{m}u_t \frac{\partial \tilde L}{\partial u}\big)
- r^{m}\pr{\rm X} \tilde L
= D_t\big( r^{m}u_t \frac{\partial \tilde L}{\partial u}-r^{m}\tilde L \big)
\end{equation}
since, by inspection, 
the Lagrangian functional 
$\tilde{\mathcal L}=\int_{t_0}^{t_1}\int_0^\infty {\tilde L}r^{m} drdt$ 
is invariant under time-translation. 
Second, we see that the dilational derivative term produces 
\begin{equation}
-u_t (rD_r +\tilde\lambda)(r^{m}u_t^2)
= -2r^{m+1} u_t^2 u_{tr} -(\tilde\lambda+m) r^{m} u_t^3
= D_r\big( -\tfrac{2}{3} r^{m+1} u_t^3 \big) + (\tfrac{1}{3} (2-m) -\tilde\lambda)r^{m} u_t^3
\end{equation}
which will reduce to a total radial derivative iff 
\begin{equation}\label{constnewcond2}
\tilde\lambda= (2-m)/3.
\end{equation}
Equating expressions \eqref{constnewcond2} and \eqref{constnew2}, 
we obtain the condition
\begin{equation}\label{pcond2}
 4k=-b(n-1)(p-2)
\end{equation}
and hence
\begin{equation}\label{scalcond2}
\lambda= -(m+1)/3,\quad m=-4(p+1)(p-2).
\end{equation}

Thus, for $a$, $b$, $k$, $\lambda$ satisfying conditions \eqref{pcond2} and \eqref{scalcond2},
the product of the variational dilation identity \eqref{morawetz2} and the symmetry characteristic function 
\eqref{symmchar2} yields a conservation law, 
where the conserved density and flux are given by $T_{11}$ and $X_{11}$ in Table~\ref{radBconslaw1st}
while the multiplier is given by $Q_{11}$ in Table~\ref{radBmultiplier1st}. 
We note that the form of this multiplier consists of $r^{m}$ times the product of 
the characteristic functions $W=ru_r-\lambda u$ and $P = u_t$ 
representing the scaling transformation and the time-translation symmetry respectively.
The similar conservation law given by 
the conserved density $T_{12}$ and flux $X_{12}$ in Table~\ref{radBconslaw1st}, 
with the multiplier given by $Q_{12}$ in Table~\ref{radBmultiplier1st}, 
can be obtained from the variational identity \eqref{morawetz2} by using 
the temporal scaling symmetry with the characteristic function $P = tu_t-u$
(in place of the previous time-translation symmetry). 

To derive the second identity, we omit the radial factor $r^{m}$ from equation \eqref{WELrad2},
which then yields
\begin{equation}\label{WELrad3}
u_{tt}-ku_{r}^pu_{rr}-(n-1)bu_{r}^{p+1}/r= E_u( L_{0}) + E_{u}(L_{1}) +\mu u_{r}^{p+1}/r
\end{equation}
where 
\begin{equation} 
 \mu=-(n-1)b
\end{equation} 
is the coefficient of the non-variational term.
Now, instead of multiplying this equation by the scaling characteristic function \eqref{BWcharfunction},
we use
\begin{equation}\label{BWcharfunction1}
 W=u_r
\end{equation}
which represents the characteristic function 
given by the generator of a translation on $r$. We thereby obtain
\begin{equation}\label{L0term3}
u_rE_u( L_0)= -E_u( {\tilde L}_0) +D_rL_0
\end{equation}
and 
\begin{equation}\label{L1term3}
u_r E_u( L_1)= -E_u( {\tilde L}_1) 
\end{equation}
with
\begin{equation}\label{L0L1new3}
 {\tilde L}_0 = \tfrac{1}{2} u_r u_{t}^{2}, \quad 
 {\tilde L}_1 =  \dfrac{k}{p+1}(2\int h(u_{r})du_{r}-u_{r}h(u_{r}))=
  \dfrac{k}{p+3}u_{r}h(u_{r})+\wt{\theta}ru_{r}
\end{equation}
where $\wt{\theta}=2k$ when $p=-2$ and $\wt{\theta}=0$ otherwise.
Substituting expressions \eqref{L0term3} and \eqref{L1term3} into equation \eqref{WELrad3} multiplied by the
characteristic function \eqref{BWcharfunction1}, we find
\begin{equation}\label{morawetz3}
u_r\big(u_{tt}-ku_{r}^pu_{rr}-(n-1)bu_{r}^{p+2}/r\big)
= -E_u({\tilde L}) -D_r(\tfrac{1}{2}u_t^2)+\mu u_{r}^{p+2}/r.
\end{equation}
Finally, we observe that the term $u_{r}^{p+2}/r$ in equation \eqref{morawetz3} will be a total 
radial derivative iff 
\begin{equation}\label{conds3}
 p=-2
\end{equation}
in which case
\begin{equation}
 \mu u_{r}^{p+2}/r=D_{r}(\mu \ln{r}).
\end{equation}
As a result, equation \eqref{morawetz3} reduces to the variational identity
\begin{equation}\label{morawetz4}
u_r\big(u_{tt}-ku_{r}^pu_{rr}-(n-1)bu_{r}^{p+2}/r\big)= -E_u({\tilde L}) -D_r(\tfrac{1}{2}u_t^2+(n-1)b\ln{r})
\end{equation}
where
\begin{equation}\label{tL2}
 \tilde L={\tilde L}_{0}+{\tilde L}_{1}
\end{equation}
is a new Lagrangian.
This identity \eqref{morawetz4} is able to produce conservation laws 
in a similar way to the previous identity \eqref{morawetz2}. 
In particular, we consider the time translation \eqref{Xtimetrans}.
Multiplying the identity \eqref{morawetz4} by the characteristic function \eqref{symmchar2}, 
we see first that the Euler-Lagrange term yields a total time derivative
\begin{equation}
-u_t E_u({\tilde L}) 
= D_t\big( u_t \frac{\partial \tilde L}{\partial u}\big)
- \pr{\rm X} \tilde L
= D_t\big( u_t \frac{\partial \tilde L}{\partial u}-\tilde L \big)
\end{equation}
since, by inspection, 
the Lagrangian functional 
$\tilde{\mathcal L}=\int_{t_0}^{t_1}\int_0^\infty {\tilde L}drdt$ 
is invariant under time-translation. 
Second, we see that the radial derivative term produces the total derivatives 
\begin{equation}
-u_t D_r (\tfrac{1}{2}u_t^2+(n-1)b\ln{r})
= D_r\big( -\tfrac{1}{6} u_t^3 \big) + D_{t}(-(n-1)bu/r).
\end{equation}

Thus, for $p=-2$, 
the product of the variational identity \eqref{morawetz4} and the symmetry characteristic function 
\eqref{symmchar2} yields a conservation law, 
where the conserved density and flux are given by $T_{10}$ and $X_{10}$ in Table~\ref{radBconslaw1st}
while the multiplier is given by $Q_{10}$ in Table~\ref{radBmultiplier1st}. 
We note that the form of this multiplier consists of just the product of 
the characteristic functions $W=u_r$ and $P = u_t$.
Interestingly, since the Lagrangian \eqref{tL2} is explicitly invariant under the shift symmetry
${\rm X}=\partial/\partial u$ with the characteristic function $P=1$, we can use the variational identity
\eqref{morawetz4} to obtain the conservation law given by $T_{7}$ and $X_{7}$ given in Table \ref{radBconslaw1st},
whose multiplier is $Q_{7}$ shown in Table \ref{radBmultiplier1st}. 

\subsection{Computational remarks.}\label{sec:conslawcomputation}

\indent\newline\indent
In the computation of conservation laws for a radial wave equation \eqref{WEreducrad}, 
the multiplier determining system \eqref{detcon1}--\eqref{detcon2} is formulated in the jet space 
whose coordinates are defined by $t$, $r$, $u$, $u_{t}$, $u_{r}$, $u_{tr}$ 
and $u_{rr}$, where $u_{tt}$ is replaced by $gu_{rr}+f$.
For both wave equations \eqref{WEreducrad1} and \eqref{WEreducrad2}, 
this determining system splits with respect to $u_{t}$, $u_{tr}$ and $u_{rr}$ 
into an overdetermined system of 13 equations for unknowns 
$Q=\alpha(t,r,u,u_{r})u_{t}+\beta(t,r,u,u_{r})$, $a$, $b$, $c$, $p$ and $n$. 
The resulting overdetermined systems are linear in the unknown multiplier components $\alpha$ and $\beta$ but are nonlinear jointly in all the unknowns. 
In particular, 
the unknown $p$ appears non-polynomially in each overdetermined system 
through respective exponents of $u$ and $u_{r}$.

We again carry out all computations by means of computer algebra.
The program {\sc Conlaw4} \cite{Wolf2002b} is used to formulate and split 
the conservation law determining system to get the overdetermined system 
which in turn is solved for the 7 unknowns 
$\alpha$, $\beta$, $a$, $b$, $c$, $p$ and $n$ by the package {\sc Crack} \cite{Wolf2002a}.
For both wave equations \eqref{WEreducrad1} and \eqref{WEreducrad2}, 
{\sc Crack} is able to complete the entire computation of multipliers automatically.
Compared to the computation of point symmetries (cf. section \ref{sec:symmcomputation}), the computational difficulties are similar 
but the total number of steps needed is approximately 2 to 4 times greater.

Once all solutions of the overdetermined system for multipliers $Q$ are obtained from {\sc Crack}, 
we compute the conserved densities $T$ and fluxes $X$
interactively in {\sc Maple 14} by evaluating the homotopy integrals \eqref{TfromQ}--\eqref{WforTX} 
whenever possible or by otherwise solving the linear PDE system \eqref{intmethodTX1}--\eqref{intmethodTX2}. 
All conditions on the parameters $a$, $b$, $c$, $p$ and $n$ for a multiplier to exist 
are imposed at the start of the {\sc Maple} computation.
 
\section{Concluding Remarks}\label{sec:concludingremarks}
In this paper we have fully classified all point symmetries \eqref{pointsymm}, kinematic conservation laws
\eqref{kinematicT}--\eqref{kinematicQ}, and generalized-energy conservation laws \eqref{genergyQ}--\eqref{energyT}
for the two classes of nonlinear radial wave equations \eqref{WEreducrad1} and \eqref{WEreducrad2} which are
parameterized in terms of the constant coefficients $a$, $b$, $c$, 
and the nonlinearity power $p$, as well as the dimension $n$.
The only restrictions placed on these five parameters are that we want each wave equation to be multi-dimensional,
hyperbolic, and semilinear or quasilinear. 

Since Noether's theorem is applicable for the cases in which these wave equations have a variational formulation,
our classification of conservation laws directly yields a corresponding classification of all variational point symmetries
and also variational contact symmetries of a certain restricted form. 
We find that such variational cases account for all of the generalized-energy conservation laws admitted by
radial wave equation \eqref{WEreducrad2} but only some of the generalized-energy conservation laws admitted by
radial wave equation \eqref{WEreducrad1}.

In the non-variational cases for radial wave equation \eqref{WEreducrad1}, we show that the additional generalized-energy
conservation laws arise through a new type of radial dilation identity that produces conservation laws from symmetries
in a different way than Noether's theorem.
A similar radial dilation identity for radial wave equation \eqref{WEreducrad2} is able to reproduce the generalized-energy
conservation laws that correspond to all of the variational contact symmetries.

Some of the symmetries and conservation laws found in our classification involve non-integer values of $n$. 
We can interpret such cases by rewriting the wave equations \eqref{WEreducrad1} and \eqref{WEreducrad2} in a 
2-dimensional form in terms of a parameter $\nu=2-n$ as follows:
\begin{subequations}\label{nuform}
\begin{equation}\label{nuform1}
 (ru_{t})_{t}=\big(r(c+bu^{p})u_{r}-\nu(cu+bH(u))\big)_{r}+(a-pb)ru^{p-1}u_{r}^{2}
\end{equation}
and
\begin{equation}\label{nuform2}
 (ru_{t})_{t}=\big(r(cu_{r}+k\wt{H}(u_{r}))-\nu cu\big)_{r}+b(1-\nu)u_{r}^{p+1}-k\wt{H}(u_{r})
\end{equation}
\end{subequations}
where
\begin{equation}
 H=\left\{\begin{aligned}&\dfrac{1}{p+1}u^{p+1},\quad&p\neq-1\\ &\ln{u},\quad&p=-1\end{aligned}\right.
\end{equation}
and
\begin{equation}
 \wt{H}=\left\{\begin{aligned}&\dfrac{1}{p+1}u_{r}^{p+1},\quad&p\neq-1\\ &\ln{u_{r}}+b(1-\nu)/k,\quad&p=-1\end{aligned}\right. .
\end{equation}

This form of the wave equations \eqref{WEreducrad1} and \eqref{WEreducrad2} is applicable for any value of $n$.
The meaning of $\nu$ can be understood through the kinematic quantity
\begin{equation}
 C(t)=\int_{0}^{\infty}urdr
\end{equation}
which satisfies
\begin{equation}\label{NSeqn}
 \dfrac{d^{2}C}{dt^{2}}=\nu N+S
\end{equation}
if $\displaystyle\lim_{r\rightarrow 0}u_{r}=0$, where
\begin{subequations}
\begin{equation}
 N=\lim_{r\rightarrow 0}(cu+bH(u)),\qquad S=\int_{0}^{\infty}(a-pb)u^{p-1}u_{r}^{2}rdr
\end{equation}
for equation \eqref{nuform1}, and
\begin{equation}
 N=\lim_{r\rightarrow 0}cu,\qquad S=\int_{0}^{\infty}b(1-\nu)u_{r}^{p+1}-k\wt{H}(u_{r})dr
\end{equation}
\end{subequations}
for equation \eqref{nuform2}.
We see from equation \eqref{NSeqn} that if $u(t,r)$ is viewed as the amplitude of a vibrating surface then
$C(t)$ describes the net transverse displacement of the surface and hence $\nu N$ has the interpretation
of a time-dependent forcing term applied to the surface at $r=0$.

The symmetries and conservation laws we have obtained in Tables \ref{radAsym} to \ref{radBgeneratorscontact} in sections 
\ref{sec:symmresults} and \ref{sec:conslawresults} can be expected to have several important applications
in the analysis of solutions for the $n$-dimensional radial wave equations \eqref{WEreducrad} 
(in the case of positive integer values of $n$) or the radial surface wave equations \eqref{nuform}
(in the case of all other values of $n$).
For example, in variational cases, both of the $n$-dimensional radial wave equations \eqref{WEreducrad} possess 
a conserved energy as well as a conserved similarity energy for $c=0$ and a conserved conformal energy
for $c=0$, $p=-4$. 
The multipliers for these energies can be used to derive useful energy identities in all non-variational cases
for these radial wave equations.
Other typical applications for our results have been mentioned in section \ref{sec:Introduction} and will be 
explained in detail elsewhere.

There are several directions in which the work in this paper can be extended. 
First, we plan to generalize our classification of symmetries and conservation laws to include
all contact symmetries \eqref{symgen} and all 1st order conservation laws \eqref{multiplierWE} for the
radial wave equations \eqref{WEreducrad} in all cases $n\neq1$.
Second, in a sequel to the present paper, we plan to determine which of the symmetries and conservation laws for the
$n$-dimensional case of the radial wave equations \eqref{WEreducrad} can be lifted to yield corresponding
symmetries and conservation laws for the translationally-invariant $n$-dimensional wave equations
\eqref{WE}.

\end{document}